\documentclass{article}
\usepackage{authblk}
\usepackage{psfrag}
\usepackage{graphicx}
\usepackage{epsfig}
\usepackage{longtable, lscape}
\usepackage{psfrag}
\usepackage{color}
\usepackage{a4wide}
\usepackage{amsmath}
\usepackage{titlesec}
\usepackage{amssymb}
\usepackage{float}
\usepackage{caption}
\usepackage{subcaption}
\usepackage{array}
\usepackage{blindtext}
\usepackage{hyperref}
\usepackage[most]{tcolorbox}
\usepackage{etoolbox} 

\begin{document}
\title{Pulsatile Magnetized $Cu$-$Al_{2}O_{3}$/Casson Blood Flow Through an Elliptical Stenotic Artery for Drug Delivery Applications}

\def\correspondingauthor{
}
\author[1]{Nimra Muqaddass} 
\author[2]{Alessandra Jannelli}
\author[3]{Francesco Oliveri}
\affil[1]{MUSAM - Multi\_scale Analysis of Materials, IMT School for Advanced Studies Lucca, Piazza S.Francesco, 19, 55100 Lucca LU, Italy;  nimra.muqaddass@imtlucca.it}
\affil[2,3]{Department of Mathematical and Computer Sciences, Physical Sciences and Earth Sciences, University of Messina, viale F. Stagno d'Alcontres 31, 98166 Messina, Italy;  alessandra.jannelli@unime.it; francesco.oliveri@unime.it}

\maketitle
\begin{abstract}
Among cardiovascular diseases, atherosclerosis is a primary cause of stenosis, involving the accumulation of plaques in the inner lining of an artery. Inspired by drug delivery applications, the proposed study aims to examine the numerical modeling of a two-dimensional, axisymmetric, and time-dependent hybrid nanofluid composed of copper $(Cu)$, alumina $(Al_{2}O_{3})$ nanoparticles, and blood as base fluid. Blood, modeled by the non-Newtonian Casson model, flows through an elliptical stenotic artery. The pulsatile nature of the pressure gradient and magnetic field impact with the Hall current parameter are also taken into account in this study. A finite difference technique, forward in time and central in space (FTCS), is deployed to numerically discretize the transformed dimensionless model using MATLAB. Comprehensive visualization of the effects of hemodynamic, geometric, and nanoscale parameters on transport characteristics, and extensive graphical results for blood flow characteristics are provided. A comparison is made among blood, regular nanofluid, and hybrid nanofluid to analyze their properties in relation to fluid flow and heat transfer. An augmentation in the non-Newtonian parameter results in an amplification of velocity and in a reduction of the temperature profile. Incorporating $Cu$ and $Al_2O_3$ nanoparticles into the fluid results in a decrease of velocity and an increase of temperature. These findings possess significant practical implications for applications where efficient heat transfer is essential, such as in drug delivery systems and the thermal management of biomedical devices. However, the observed reduction in velocity may necessitate modifications to flow conditions to ensure optimal operational performance in these contexts.

\end{abstract}
\section{Introduction}
\begin{figure}[H]
\centering
\includegraphics[width=0.7\linewidth]{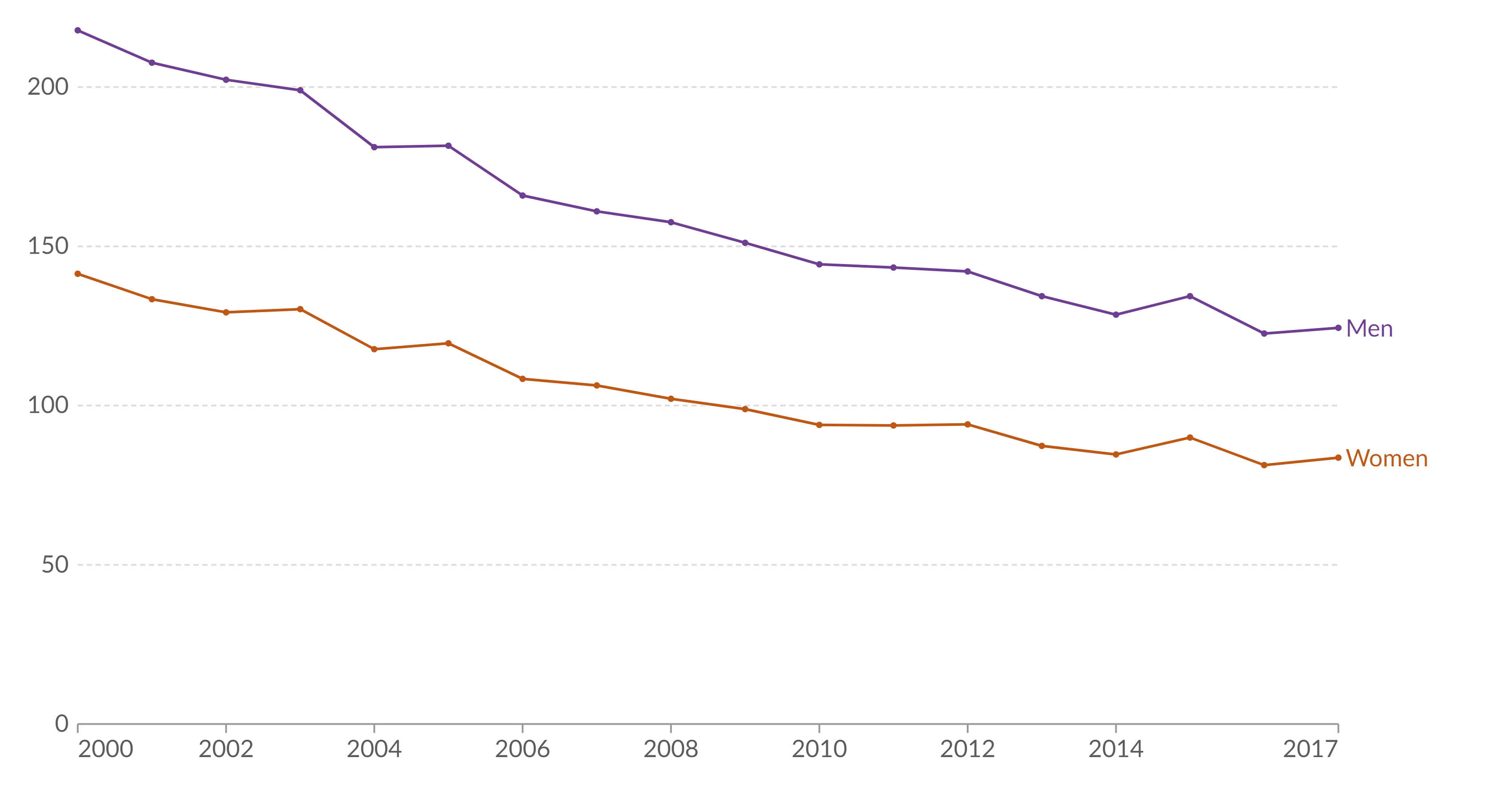}
\caption{The number of deaths caused by CVDs in Italy, categorized by gender and on a yearly basis, from 2000 to 2017 (\emph{Our World in Data}, \url{https://ourworldindata.org/} \cite{Hannah}).}
\label{cvdItaly}
\end{figure}

\begin{figure}[h]
\centering
\begin{subfigure}{0.5\textwidth}
\includegraphics[width=0.8\linewidth, height=5cm]{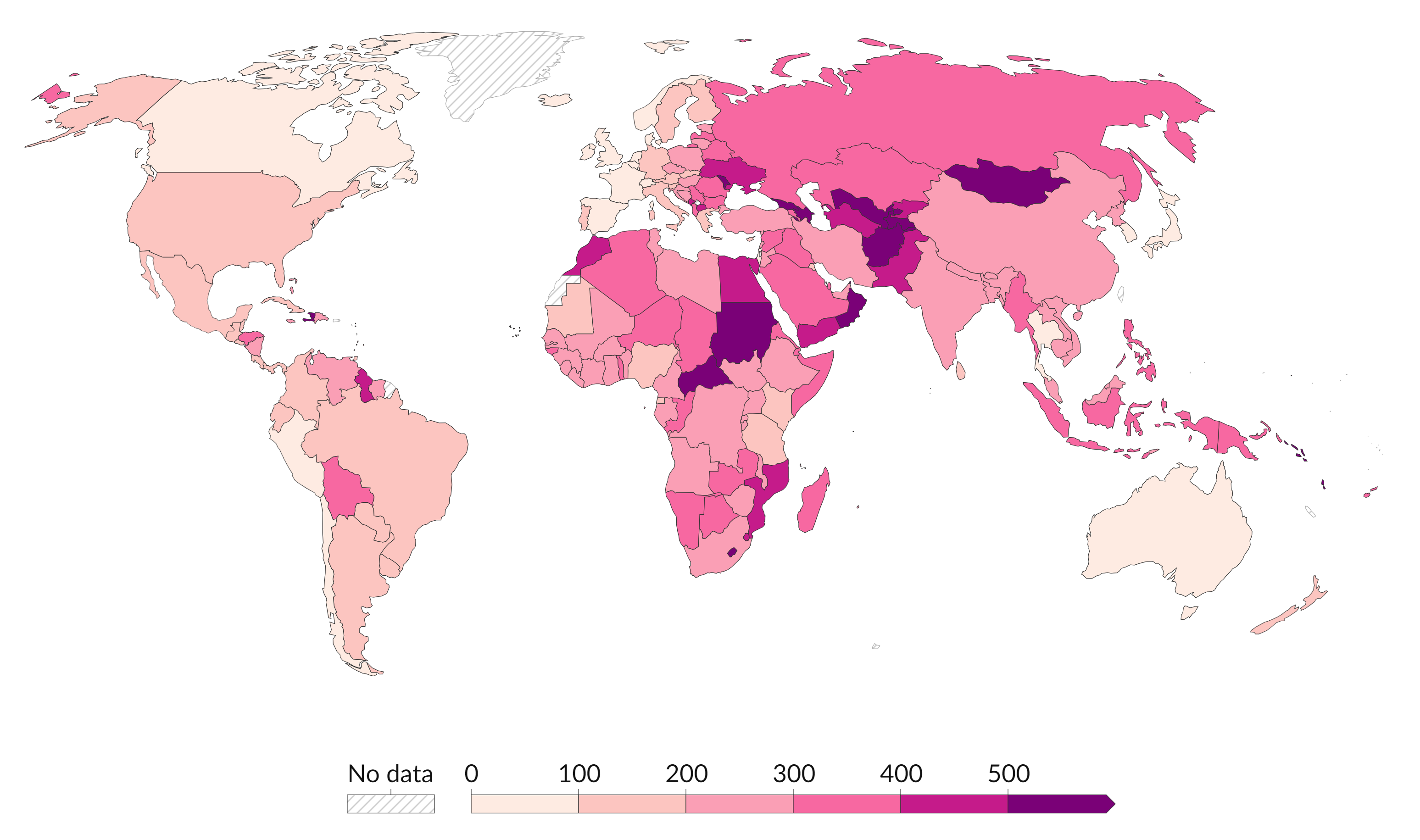} 
\caption{}
\end{subfigure}\begin{subfigure}{0.5\textwidth}
\includegraphics[width=0.8\linewidth, height=5cm]{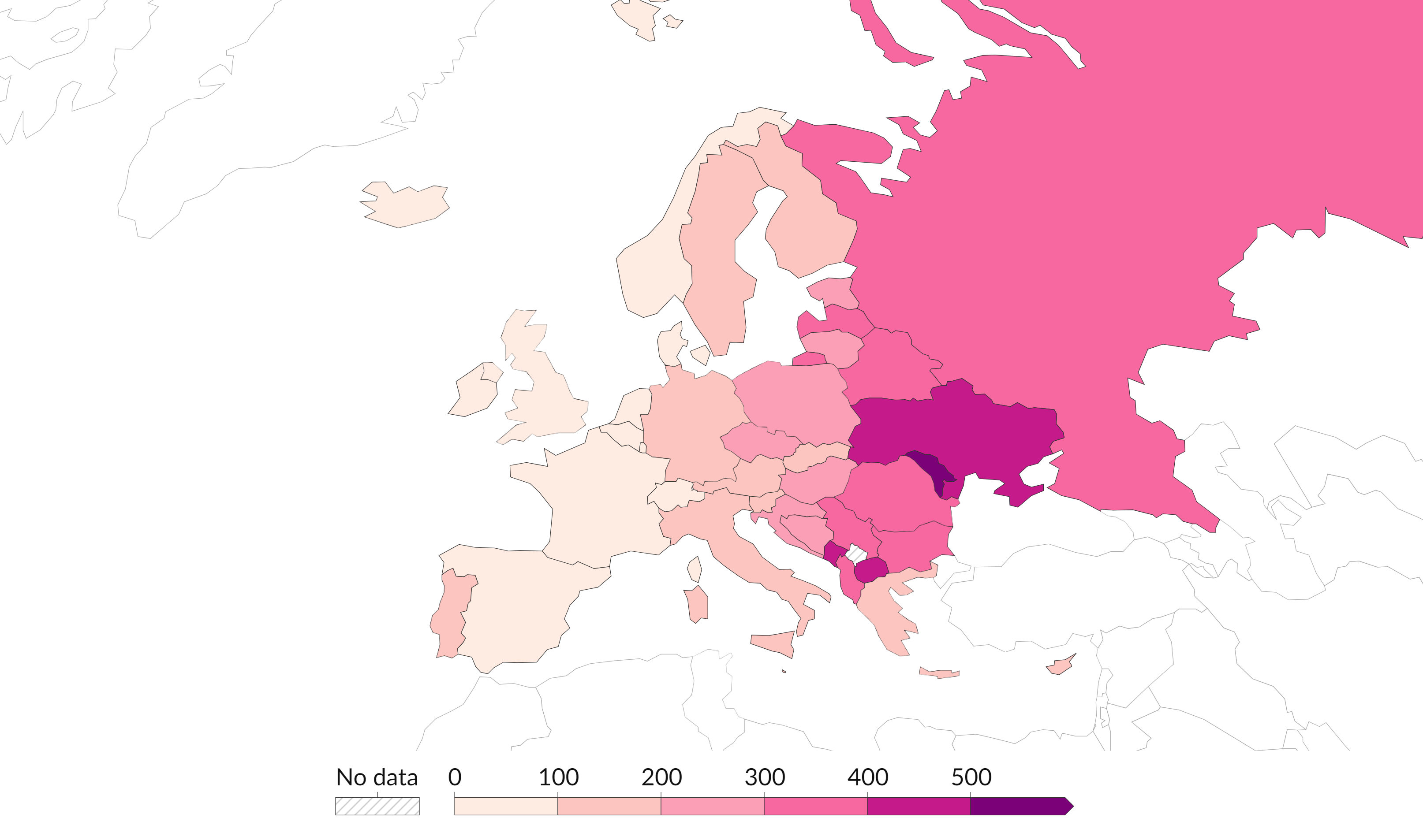}
\caption{}
\end{subfigure}
\caption{\textbf{(a)} Globally, and per 100,000 population, the average death rate from CVDs in 2017. \textbf{(b)} Europe's 2017 CVD death rate: per 100,000. The data and images were adapted from \emph{Our World in Data} (\url{https://ourworldindata.org/} \cite{Hannah}).}
\label{dataCVD}
\end{figure}

Cardiovascular diseases (CVDs) remain a global health concern, necessitating continual advancements in the understanding and management of arterial conditions. Throughout a person's lifetime, the heart pumps approximately five liters of blood into the cardiovascular circulatory system every minute in order to maintain the proper functioning of organs, tissues, and cells. Although this system has evolved to be highly effective, it is still vulnerable to various heart and vascular disorders. In fact, CVDs are today the most common cause of death both in Italy and worldwide \cite{Hannah} (Figure \ref{cvdItaly} and Figure \ref{dataCVD}). According to the World Health Organization (WHO) \cite{Shanthi}, many premature deaths, including three million people under sixty years old, could have been prevented through improved diagnostics and interventions. For proper function and survival, the cardiovascular system, which includes the heart, blood vessels, and blood, is responsible for a continuous supply of nutrients and oxygen, as well as the removal of waste products \cite{Paulsen, MONKHOUSE}.\\
The presence of red blood cells plays a vital role in determining the behavior of blood and its properties 
(see Table~\ref{crouch}). Therefore, it is crucial to examine hemodynamic factors related to blood and vessels to gain a better understanding of arterial diseases \cite{Reference0a4c1}. Numerous research studies have showcased the successful application of experimental and numerical methods in the field of medical treatment. Among these, in \cite{Reference0a5c1} an investigation into the wall shear stress of blood, utilizing the Carreau-Yasuda fluid model, has been carried out. Concurrently, in  \cite{Reference0a6c1} the blood has been assumed as a Newtonian fluid, comparing their velocity profile and wall shear stress distribution findings to those of \cite{Reference0a7c1}. 
In \cite{YAN2020105434}, it has been investigated the hemodynamic rheology inside an arterial segment having a cone shape of stenosis.

\begin{table}[H]
\begin{center}
\caption{An overview of blood components
\cite{Reference14c1}.}
\label{crouch}
\begin{tabular}{ | m{4cm} | m{4cm}| m{2.5cm} |} 
\hline
\textbf{Blood Components} & \textbf{Principle Functions} & \textbf{\% by Volume} \\ 
\hline
Plasma & Transports blood, carbon dioxide, and nutrients & 50–60 \\ 
\hline
Red cells & Oxygen carriers & 40–50 \\ 
\hline
White cells & Body's immune system & 0.7 \\ 
\hline
Platelets & Clot formation of blood & 0.3 \\ 
\hline
Macromolecules: albumin & To maintain oncotic pressure & 02 \\ 
\hline
Other & Various & 1.5 \\ 
\hline
\end{tabular}
\end{center}
\end{table}
In \cite{Reference0ac1}, the blood flow through the cardiovascular system, which is one of the essential phenomena in biomedical engineering, has been studied. For doctors, the turbulent blood flow is more interesting, which might happen due to the increase in flow to a particular value, narrowing of the blood vessels, or the presence of eddies and stenosis swirling vortexes. In \cite{Reference0bc1}, the blood has been considered as a homogeneous fluid with different properties relative to the size of the blood vessel, whereas in \cite{ELHANAFY2020113550}  the hemodynamic characteristics of blood flow in arterial stenotic sections with multiple degrees of stenosis has been numerically evaluated. Transport of the molecules in a fluid is governed by convective-diffusive laws, and in \cite{Reference0cc1}, using the Navier–Stokes equations, the blood has been considered as a viscous, incompressible, and homogenous fluid. Moreover, in \cite{HALDAR}, it has been analyzed the blood flow in stenotic regions of various shapes and identified how these areas can impede the flow in arterial segments. The blood flow problem for an arterial stenotic region was mathematically modeled in \cite{CHAKRAVARTY}. Within a tapered stenotic region, in \cite{MANDAL2005151} the non-Newtonian blood flow modeled with a power law has been investigated numerically for time variable stenosis.  Several computational approaches have been employed to study blood flow in stenosed arteries. Finite element methods (FEM), finite difference methods (FDM), and finite volume methods (FVM) are commonly used numerical techniques to solve the governing equations of fluid dynamics, such as the Navier-Stokes equations, in complex geometries typical of vascular structures. These simulations allow researchers to investigate parameters such as velocity profiles, wall shear stress, pressure gradients, and flow patterns, which are critical for understanding the biomechanical forces acting on arterial walls and their implications for disease progression \cite{cimmelli2011thermodynamics, Supratim, Ha, Sohail, Carvalho, Sarwar, Vasu, Koyel, cavaccini2006mathematical}. \\
After a deep analysis of previously published data, we have furnished comprehensive momentum and energy profiles, along with mass and heat transfer analysis, for the non-Newtonian hybrid nano-blood flow through an elliptical-shaped stenotic arterial segment. The set of partial differential equations, accounting for the aforementioned considerations, has been written in dimensionless form and subsequently solved numerically using the forward time central space differentiation method. The resulting numerical simulations have been compared with existing literature and are presented graphically for thorough evaluation.  
\section{Problem Formulation} \label{sec2}
\begin{figure}[h]
\centering
\includegraphics[width=12cm, height=7.5cm]{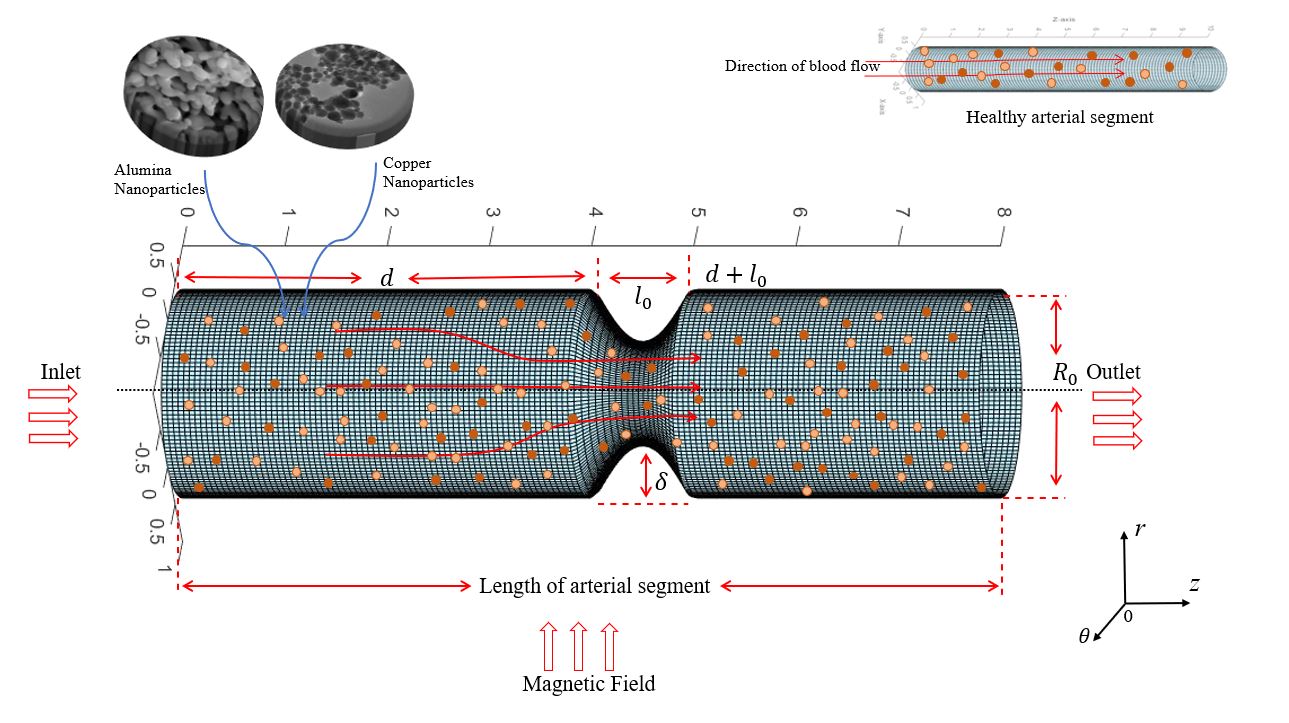}
\caption{The Schematic representation of the problem.}
\label{flowDiagram}
\end{figure}

It is usually assumed that blood flow is laminar and Newtonian, but it behaves as non-Newtonian fluid when flowing through a stenotic artery. Let us consider a time-dependent, laminar, and incompressible blood flow modeled by a non-Newtonian Casson fluid model introduced by British physicist and rheologist Roger Casson \cite{casson1959} in 1959 (equation \eqref{CassonModel}) with velocity components $(\bar u(\bar r,\bar z,\bar t),\, \allowbreak\bar w (\bar r,\bar z,\bar t))$ flowing through an elliptical-shaped stenotic artery in cylindrical coordinates $(\bar r,\bar z)$ as shown in Figure \ref{flowDiagram}. Moreover, a uniform magnetic field $\beta_{0}$ is applied along the radial direction. The model takes the Hall current into account, which generates and affects the flow of electricity under a stronger magnetic field. Moreover, suspension of copper $Cu$ and aluminum oxide $Al_{2}O_{3}$ nanoparticles are also introduced in the blood (base fluid), forming $Cu$-$Al_{2}O_{3}$-blood hybrid nanofluid. The region for axisymmetric stenosis is described by the function in equation (\ref{stenosisGeometry}), where the arterial circular cross-section is $R_{0}$  except the region of stenosis, $L$ is the length of total arterial section considered, $\bar R(\bar z)$ is the radius of the stenotic region, $\delta$ is the stenosis intensity and $l_{0}$ is the length of stenosis. 

The Casson model can be expressed through the equation
\begin{equation}
\label{CassonModel}
\tau_{ij} =\begin{cases}
2 \bigg( \mu_{b} + \frac{\textit{P}_{y} }{\sqrt{2  \pi}} \bigg) e_{ij}, & \text{ $\pi > \pi_c $},\\
2 \bigg( \mu_b + \frac{\textit{P}_{y} }{\sqrt{2  \pi _c}} \bigg) e_{ij}, & \text{ $\pi \le \pi_{c} $}, 
\end{cases}
\end{equation}
where $\tau_{ij}$ is shear stress tensor, $\mu_{b}$ is the Casson viscosity of non-Newtonian fluid, $P_{y} = \mu_b \sqrt{2\pi}/\beta$ is the yield stress of fluid, $\pi = e_{ij} \cdot e_{ij}$ is the product of deformation rate to itself, $\pi_c$ is the critical value of $\pi$ according to the non-Newtonian model, $e_{ij}$ is strain rate and $\mu_{c} $ is the critical value based on non-Newtonian model. For $\pi\le\pi_{c}$ with $\beta = \frac{ \bar \mu_{b} \sqrt{2 \bar \pi _{c}} }{\Bar{\textit{P}}_{y}}$ as Casson model parameter, the shear stress tensor becomes: 
\begin{equation*}
\tau_{ij} = 2  \mu_{b} \bigg(1+ \frac{1}{\beta} \bigg)e_{ij}
\end{equation*}
An elliptical artery lesion can be modeled by modifying the radius of a healthy artery, $R_{0}$, along the axial direction $\bar z$ to simulate narrowing due to stenosis. The mathematical equation to describe a partially obstructed artery is as follows:
\begin{equation}
\label{stenosisGeometry}
\bar R(\bar z) =\begin{cases}
R_{0} - \delta^{*} sin \bigg( \pi \bigg( \frac{\bar z - \bar d}{l_{0}}\bigg) \bigg), & \text{ $\bar d \le \bar z \le \bar d+l_{0}$}  \\
R_{0} &  \hbox{otherwise}. 
\end{cases}
\end{equation}
The variation in blood pressure along the vessel length reflects in a consequent pressure gradient within a human artery. Aorta arteries have the highest blood pressure due to direct heart pumping. As blood moves away from the heart, the pressure gradually decreases. 

To do the simulation of realistic blood flow dynamics and model the oscillatory nature of blood pressure in arteries due to the heartbeat, the pressure gradient along the axial direction $\bar z$ is assumed to be modeled by
\begin{equation}
\label{pGrad}
-\frac{\partial \bar p}{\partial \bar z} = p_{0} + p_{1} \cos{\omega \bar t}, \quad  t>0, 
\end{equation}
where $p_{0}$ is the pressure gradient amplitude representing the steady-state component of the pressure gradient due to the continuous flow of blood. A typical value of $p_{0}$ might range from 1-5 $(mm\,Hg/cm)$; $p_{1}$ is the amplitude of the heart's cyclic contractions and relaxations components and can be around 5-20 $mm\,Hg$; moreover, $\omega =2\pi f_{p}$, where $f_{p}$ is the pulse frequency, models the pulsatile nature of blood flow.
Based on the assumptions made above, the dimensional governing flow and heat equations can be expressed as follows:
\begin{align}
\label{c2-c}
&\frac{{\bar u}}{{\bar r}} + \frac{\partial {\bar u}}{\partial {\bar r}} + \frac{\partial {\bar w}}{\partial {\bar z}} = 0,\\
&\rho_{hnf} \bigg( \frac{\partial { \bar u}}{\partial {\bar t}} + \bar{u} \frac{\partial {\bar u}}{\partial {\bar r}} + \bar w \frac{\partial {\bar u}}{\partial {\bar z}} \bigg) = -\frac{\partial \bar p }{\partial \bar r} + \frac{ \sigma_{hnf} \beta_{0}^{2}}{(m^2 +1)} \big(\bar u + m \bar w \big)\nonumber\\
&\qquad+ \mu_{hnf} \bigg( 1+\frac{1}{\beta} \bigg) 
\bigg[ \frac{\partial^{2}{\bar u}}{\partial {\bar r}^{2}} + \frac{1}{\bar r} \frac{\partial \bar u}{\partial \bar r} + \frac{\partial^{2}{\bar u}}{\partial {\bar z}^{2}} - \frac{\bar u}{\bar r^{2}} \bigg],
\label{c2-v_mag}\\
&\rho_{hnf} \bigg( \frac{\partial { \bar w}}{\partial {\bar t}} + \bar{u} \frac{\partial {\bar w}}{\partial {\bar r}} + \bar w \frac{\partial {\bar w}}{\partial {\bar z}} \bigg) =  p_{0} + p_{1} \cos{\omega \bar t} + \frac{\sigma_{hnf} \beta_{0}^{2}}{(m^2 +1)} \big(\bar w - m \bar u \big)\nonumber\\
\label{c2-w_mag}
&\qquad+ \mu_{hnf} \bigg( 1+\frac{1}{\beta} \bigg) 
\bigg[ \frac{\partial^{2}{\bar w}}{\partial {\bar r}^{2}} + \frac{1}{\bar r} \frac{\partial \bar w}{\partial \bar r} + \frac{\partial^{2}{\bar w}}{\partial {\bar z}^{2}} \bigg] ,
\end{align}
and energy equation becomes:
\begin{align}
&(\rho C_p)_{hnf} \bigg( \frac{\partial \bar T}{\partial \bar t} + \bar u \frac{\partial \bar T}{\partial \bar r} + \bar w \frac{\partial \bar T}{\partial \bar z} \bigg) = K_{hnf} \bigg[\frac{\partial^2\bar T}{\partial \bar r^2} + \frac{1}{\bar r}\frac{\partial \bar T}{\partial \bar r}+\frac{\partial^2\bar T}{\partial \bar z^2} \bigg] \nonumber\\
&\qquad+2\mu_{hnf}\bigg(1+\frac{1}{\beta}\bigg)\bigg[\bigg(\frac{\partial \bar u}{\partial \bar r} \bigg)^2 + \bigg(\frac{\bar u}{\bar r} \bigg)^2 +\bigg(\frac{\partial \bar w}{\partial \bar z} \bigg)^2 + \frac{1}{2}\bigg(\frac{\partial \bar u}{\partial \bar z} +\frac{\partial \bar w}{\partial \bar r}\bigg)^2 \bigg]\nonumber\\
\label{c2-t_mag}
&\qquad +\frac{\sigma_{hnf} \beta_0^2}{(m^2 +1)} \big(\bar w - m \bar u \big)^2
\end{align}
In the above equations,$\bar u$, $\bar w)$, $\bar t$, $\bar T$, $\bar p$ are correspondingly dimensional velocities along radial and axial directions, time, temperature, and pressure; moreover, $\rho_{hnf}$, $\mu_{hnf}$, $\sigma_{hnf}$, $(C_{p})_{hnf}$ and $\kappa_{hnf}$ are density, dynamic viscosity, electrical conductivity, specific heat capacity and thermal conductivity for hybrid nanofluids respectively. 
 
The equations are solved with suitable initial and boundary conditions, say
\begin{equation}
\label{c2-bc}
\begin{aligned}
& \bar{w} = 0, \,\,\,\,\,\ \bar{T} = 0,  \quad \hbox{at} \quad \bar{t}=0,\\
&  \frac{\partial {\bar w}}{\partial {\bar r}} = 0, \qquad  \frac{\partial {\bar T}}{\partial {\bar r}} = 0,  \quad \hbox{at}\quad  \bar{r}=0,\\
&\bar{w} = 0, \qquad \bar{T} = \bar {T}_{w},  \quad   \hbox{at} \quad \bar{r}=\bar{R},
\end{aligned}
\end{equation}
\emph{i.e.}, we assume that the blood is initially at rest within the stenosed artery, and a reference temperature  normalized to zero. Also,  there is no radial change in velocity or temperature, implying a symmetric flow, which is characteristic of axisymmetric flow scenarios. Finally, at the arterial wall $\bar r = \bar R$, the no-slip condition is applied where the fluid velocity matches that of the solid boundary, and the temperature $\bar T$ is set to the arterial wall temperature $\bar T_{w}$.

The governing nonlinear equations and the initial and boundary conditions need to be transformed via a suitable set of dimensionless variables, say
\begin{equation}
\label{c2-nd}
\begin{aligned}
& r=\frac{\bar r}{R_{0}},\,\,\ z=\frac{\bar z}{l_{0}},\,\,\ t=\frac{u_{0}\bar t}{R_{0}},\,\,\ R=\frac{\bar R}{R_{0}},\,\,\ d=\frac{\bar d}{l_{0}},\,\,\ u=\frac{l_{0}\bar u}{\delta^* u_{0}},\,\,\ w=\frac{\bar w}{u_{0}},\\
& \theta=\frac{\bar T - \bar T_{1}}{\bar T_{w} - \bar T_{1}}, \,\,\ p=\frac{R_{0}^2\bar p}{u_{0}l_{0}\mu_{0}}, \,\,\ Re=\frac{R_{0}\rho_{f}u_{0}}{\mu_{f}}, \,\,\  Pr=\frac{c_{p}\mu_{f}}{\kappa_{f}}, \,\,\ \delta=\frac{\delta*}{R_{0}}, \,\,\ \epsilon=\frac{R_{0}}{l_{0}},\\
&M^2=\frac{\sigma_{f} R_{0}^2\beta_{0}^2}{\mu_{f}}, \,\,\ Ec=\frac{u_{0}^2}{C_{p}(\bar T_{w}-\bar T_{1})},
\end{aligned}
\end{equation}
where $u_{0}$ is the reference velocity, $T_{w}$ the wall temperature, $l_{0}$ the stenosis length, $R_{0}$ the radius of artery, $\epsilon$ the vessel aspect ratio, and $\delta$ the stenosis height; moreover, $Re$, $M$, $Pr$, and $Ec$ are dimensionless Reynolds number, Hartmann number, Prandtl number, and Eckert number, respectively.

This approach facilitates comparison of solutions among different systems. It identifies dimensionless groups, like Reynolds and Prandtl numbers, which reveal the essence of the flow and the relative importance of different forces. The long wavelength approximation in fluid mechanics often involves dimensionless parameters, such as $\delta << 1$ and $\epsilon = O(1)$, representing characteristic length scale and scaling parameter, which implies that one length scale is much smaller than another and certain terms in the equations are of the same order of magnitude as $\epsilon$ and thus relatively small. 

Therefore, equations (\ref{c2-c})-(\ref{c2-bc}) assume the form
\begin{align}
\label{c2-ndc1}
&\frac{\partial w}{\partial z} = 0,\\
\label{c2-ndv1}
&\frac{\partial p}{\partial r} = 0,\\
 &\frac{\partial w}{\partial t}= \frac{\rho_{f}}{\rho_{hnf}}\frac{1}{Re}\bigg[A_{0}\big(1+e \cos(c_{1}t)\big) + \frac{\mu_{hnf}}{\mu_{f}}\bigg(1+\frac{1}{\beta} \bigg) \bigg(\frac{\partial^2 w}{\partial r^2} +\frac{1}{r}\frac{\partial w}{\partial r} \bigg) 
\nonumber\\
\label{c2-ndw1}
&\qquad+ \frac{M^2}{(m^2+1)} \frac{ \sigma_{hnf}}{\sigma_{f}} w\bigg],\\
\label{c2-ndt1}
&\frac{\partial \theta}{\partial t}= \frac{(\rho C_{p})_{f}}{(\rho C_{p})_{hnf}}\bigg[\frac{\kappa_{hnf}}{\kappa_{f}}\frac{1}{Re Pr} \bigg(\frac{\partial^2 \theta}{\partial r^2} + \frac{1}{r}\frac{\partial \theta}{\partial r} \bigg) + \bigg(1+\frac{1}{\beta} \bigg)
\frac{\mu_{hnf}}{\mu_{f}}\frac{Ec}{Re} \bigg(\frac{\partial w}{\partial r} \bigg)^2 \bigg],
\end{align}
and the initial/boundary conditions are as follows:
\begin{equation}
\label{c2-ndbc1}
\begin{aligned}
& w = 0, \qquad \theta = 0,  \quad \hbox{at} \quad t=0,\\
&  \frac{\partial w}{\partial r} = 0, \,\,\,\  \frac{\partial \theta}{\partial r} = 0,  \quad \hbox{at} \quad r=0,\\
&w = 0, \qquad \theta = 1,  \quad   \hbox{at} \quad r=R.
\end{aligned}
\end{equation}
The set of equations (\ref{c2-ndw1})-(\ref{c2-ndbc1}) are further transformed setting $(x = r/R(z))$, so that we obtain a more concise and simplified set of equations, namely: 
\begin{align}  
&\frac{\partial w}{\partial t}= \frac{\rho_{f}}{\rho_{hnf}}\frac{1}{Re}\bigg[A_{0}\big(1+e \cos(c_{1}t)\big) + \frac{\mu_{hnf}}{\mu_{f}}\frac{1}{R^2}\bigg(1+\frac{1}{\beta} \bigg) \bigg(\frac{\partial^2 w}{\partial x^2} +\frac{1}{x}\frac{\partial w}{\partial x} \bigg) \nonumber\\
\label{c2-w1}
&\qquad+ \frac{M^2}{(m^2+1)} \frac{ \sigma_{hnf}}{\sigma_{f}} w\bigg],\\
\label{c2-t1}
&\frac{\partial \theta}{\partial t}= \frac{(\rho C_{p})_{f}}{(\rho C_{p})_{hnf}}\bigg[\frac{\kappa_{hnf}}{\kappa_{f}}\frac{1}{Re Pr} \frac{1}{R^2}\bigg(\frac{\partial^2 \theta}{\partial x^2} + \frac{1}{x}\frac{\partial \theta}{\partial x} \bigg) + \bigg(1+\frac{1}{\beta} \bigg)
\frac{\mu_{hnf}}{\mu_{f}}\frac{Ec}{Re} \frac{1}{R^2} \bigg(\frac{\partial w}{\partial x} \bigg)^2 \bigg],
\end{align}
and the associated initial/boundary conditions write:
\begin{equation}
\label{c2-bc1}
\begin{aligned}
& w(x,0) = 0, \qquad \theta(x,0) = 0,  \\
&  \frac{\partial w(0,t)}{\partial x} = 0, \,\,\,\  \frac{\partial \theta (0,t)}{\partial x} = 0,  \\
&w(1,t) = 0, \qquad \theta (1,t) = 1.
\end{aligned}
\end{equation}
The dimensionless form of equation (\ref{stenosisGeometry}), illustrating the elliptical stenosis for the present analysis,
is: 
\begin{equation}
\label{stenosisGeometrynd}
R(z) =\begin{cases}
1 - \delta \sin \big( \pi \big(z-d\big) \big), &d \le  z \le  d+1,  \\
1 &  \hbox{otherwise}. 
\end{cases}
\end{equation}
Copper $(Cu)$ and alumina $(Al_2O_3)$ nanoparticles show unique properties that make them promising candidates for drug delivery in medicine. These nanoparticles are highly versatile in drug delivery applications due to their adjustable surface properties and high surface area-to-volume ratios, allowing for precise dosage regulation and enhanced therapeutic efficacy. The concentration of nanoparticles in drug delivery is a critical factor that determines the effectiveness and safety of the delivery system. Typically, the volume fraction of copper nanoparticles falls within the lower end of the spectrum, ranging from 0.1\% to 5\%, and for $Al_2O_3$, the volume fraction commonly falls within the range of 0.1\% to 10\%. The thermophysical features, i.e., density, thermal conductivity, electrical conductivity, and specific heat for base fluid (blood), copper $(Cu)$ and aluminum oxide $(Al_{2}O_{3})$  are given in the Table \ref{table:1} and the relation between both nanofluids are mentioned in Table \ref{table:2} and Table \ref{table:3}. $\mu_{f}$, $\sigma_{f}$, $(C_{p})_{f}$, $\rho_{f}$ and $\kappa_{f}$ are viscosity, electrical conductivity, specific heat capacity, density and thermal conductivity of the base fluid. $(\phi_{1},\phi_{2})$ are nanoparticle volume fraction. $s_{1}$, $s_{2}$ and $b_f$ denotes first, second particles and base fluid. 
\begin{table}[H]
\caption{Thermophysical features for $Cu$, $Al_{2}O_{3}$ and blood \cite{Reference6c2}.} 
\label{table:1}
\begin{center}
\begin{tabular}{|c c c c|} 
\hline
Thermophysical Properties & $Cu$ & $Al_{2}O_{3}$ & Human Blood \\ [0.5ex] 
\hline
Density $kg m^{-3}$ & 8933 & 3970 & 1063 \\
Specific thermal capacity $J Kg^{-1}K^{-1}$ & 385 & 765 & 3594 \\ 	
Thermal conductivity $W m^{-1} K^{-1}$ & 401 & 40 & 0.492 \\			
Electrical conductivity $\Omega m^{-1}$  & $59.6 \times 10^{6}$ & $35 \times 10^{6}$ & $6.67 \times 10^{-1}$ \\
\hline
\end{tabular}
\end{center}
\end{table}
\begin{table}[H]
\caption{properties of nanofluid \cite{Reference6c2}.} 
\label{table:2}
\begin{center}
\begin{tabular}{|c c|} 
\hline
Properties & Nanofluid \\ [0.5ex] 
\hline
 Dynamic viscosity & $\mu_{nf} = \frac{\mu_{f}}{(1-\phi)^{2.5}}$ \\
 Density & $\rho_{nf} = (1-\phi)\rho_{f} - \phi\rho_{s} $ \\
 Thermal conductivity & $\kappa_{nf} = \kappa_{f}\bigg(\frac{\kappa_{s}+2\kappa_{f}-2\phi(\kappa_{f}-\kappa_{s})} {\kappa_{s}+2\kappa_{f}+\phi(\kappa_{f}-\kappa_{s)}}\bigg) $ \\ 
 Electrical conductivity & $\sigma_{nf} = \sigma_{f}\bigg(\frac{2\sigma_{f}+\sigma_{s}-2\phi(\sigma_{f}-\sigma_{s})}{2\sigma_{f}+\sigma_{s}+\phi(\sigma_{f}-\sigma_{s})}\bigg)$  \\
Heat capacity & $(\rho C_{p})_{nf} = (1-\phi)(\rho C_{p})_{f}-\phi(\rho C_{p})_{s}$ \\
\hline
\end{tabular}
\end{center}
\end{table}
\begin{table}[h]
\caption{Thermo-physical properties of hybrid nanofluid \cite{Reference6c2},\cite{Reference7c2},\cite{Reference8c2},\cite{Reference9c2},\cite{Reference10c2}.} 
\label{table:3}
 \begin{center}
\begin{tabular}{|c c|} 
\hline
Properties & Hybrid nanofluid \\ [0.5ex] 
\hline 
Dynamic viscosity & $\mu_{hnf} = \frac{\mu_{f}}{(1-\phi_{1})^{2.5}(1-\phi_2)^{2.5}}$ \\
Density & $\rho_{hnf} = [(1-\phi_{2})((1-\phi_{1})\rho_{f}+\phi_{1}\rho_{1})] + \phi_{2}\rho_{2} $ \\
 Thermal conductivity & $\frac{\kappa_{hnf}}{\kappa_{bf}} = \bigg(\frac{(\kappa_{s2}+2\kappa_{bf})-2\phi_{2}(\kappa_{bf}-\kappa_{s2})} {(\kappa_{s2}+2\kappa_{bf})+\phi_{2}(\kappa_{bf}-\kappa_{s2)}}\bigg) $ ,\\
& $\frac{\kappa_{bf}}{\kappa_{f}} = \bigg(\frac{(\kappa_{s1}+2\kappa_{f})-2\phi_{1}(\kappa_{f}-\kappa_{s1})} {(\kappa_{s1}+2\kappa_{bf})+\phi_{1}(\kappa_{f}-\kappa_{s1)}}\bigg) $ \\
Electrical conductivity & $\sigma_{hnf} = \sigma_{bf}\bigg(\frac{2\sigma_{bf}+\sigma_{s2}-2\phi_{2}(\sigma_{bf}-\sigma_{s2})}{2\sigma_{bf}+\sigma_{s2}+\phi_{2}(\sigma_{bf}-\sigma_{s2})}\bigg)$ , \\
&   $\sigma_{bf} = \sigma_{f}\bigg(\frac{2\sigma_{f}+\sigma_{s1}-2\phi_{1}(\sigma_{f}-\sigma_{s1})}{2\sigma_{f}+\sigma_{s1}+\phi_{1}(\sigma_{f}-\sigma_{s1})}\bigg)$    \\
Heat capacity & $(\rho C_{p})_{hnf} = (\rho C_{p})_{f}(1-\phi_{2})\bigg( (1-\phi_{1})+\phi_{1}\frac{(\rho C_{p})_{s1}}{(\rho C_{p})_{f}} \bigg) + \phi_{2}(\rho C_{p})_{s2}$ \\
\hline
\end{tabular}
\end{center}
\end{table}
Local Nusselt number and wall shear stress are the two important parameters that are often studied in relation to blood flow dynamics. Mathematical expressions for wall shear stress and local Nusselt number are:
\begin{equation}
\label{c2-nu}
\tau_{w} = - \mu_{hnf} \bigg(\frac{\partial w}{ \partial r} \bigg)_{r=R(z)}, \,\,\,\,\,\,\,\,\,\,\  Nu_{l} = - \bigg(\frac{\partial \theta}{ \partial r} \bigg)_{r=R(z)},
\end{equation}
whence, using the radial coordinate transformation, write as:
\begin{equation}
\label{c22-ndnu}
\tau_{w} = - \frac{\mu_{hnf}}{R(z)}  \bigg(\frac{\partial w}{ \partial x} \bigg)_{x=1}, \,\,\,\,\,\,\,\,\,\,\ Nu_{l} = - \frac{1}{R(z)}\bigg(\frac{\partial \theta}{ \partial x} \bigg)_{x=1}.
\end{equation}
\section{Numerical methods}
The dimensionless IBVP in equations \eqref{c2-w1}--\eqref{stenosisGeometrynd} and related flow properties in equations \eqref{c22-ndnu} are simulated numerically via explicit forward time central space (FTCS) method, which approximates the derivative of functions by discretization of the continuous domain in both space and time into a mesh or grid. $\Delta x$ and $\Delta z$ represent the distance between adjacent points in 2D space, determining the resolution of the simulation. Meanwhile, $\Delta t$ represents the distance between adjacent time steps, which determines the accuracy of the simulation over time. In the intervals $0 \le x \le xmax$, $0 \le z \le zmax$ and $0 \le t \le tmax$, we have
\begin{align*}
& x_{j} =(j-1) \Delta x, \qquad j=1,2,\ldots,J,\\
&z_{m} = (m-1)\Delta z, \qquad m=1,2,\ldots,M,\\
& t_{n} = (n-1)\Delta t, \qquad n=1,2,\ldots,N,
\end{align*}
where $J$, $M$ and $N$ are the total number of space and time steps or \textit{nodes} including boundary. Size of time step $\Delta t$ and space steps $\Delta x$, $\Delta z$ can be expressed as follows:
\begin{align*}
    \Delta x =  \frac{xmax}{J-1}, \,\,\,\,\,\  \Delta z =  \frac{zmax}{M-1}, \,\,\,\,\,\  \Delta t =  \frac{tmax}{N-1}.
\end{align*}
The discretized form of the current problem takes the following form:
\begin{align}
w(j,n+1)&= w(j,n)+\Delta t \frac{\rho_{f}}{\rho_{hnf} Re}\bigg[A_{0}\big(1+e \cos(c_{1}t(n)\big) + \frac{\mu_{hnf}}{\mu_{f}}\frac{1}{(R(m))^2}\bigg(1+\frac{1}{\beta} \bigg) \times \nonumber \\
&\times \bigg(\frac{w(j+1,n)-2w(j,n)+w(j-1,n)}{\Delta x^2} + \frac{1}{x(j)} \frac{w(j+1,n)-w(j-1,n)}{2 \Delta x} \bigg) \nonumber \\
 &+\frac{M^{2}}{m^{2}+1} \frac{ \sigma_{hnf}}{\sigma_{f}} w(j,n) \bigg],\label{c2-ftcsw}\\
\theta(j,n+1) &= \theta(j,n) + \Delta t \frac{(\rho C_{p})_{f}}{(\rho C_{p})_{hnf}}\bigg[\frac{\kappa_{hnf}}{\kappa_{f}}\frac{1}{Re Pr} \frac{1}{(R(m))^2}\bigg( \frac{\theta(j+1,n)-2\theta(j,n)+\theta(j-1,n)}{\Delta x^2}\bigg) \\  
&+\frac{1}{x(j)} \frac{\theta(j+1,n)-\theta(j-1,n)}{2 \Delta x} + \bigg(1+\frac{1}{\beta} \bigg) \frac{\mu_{hnf}}{\mu_{f}} \frac{Ec}{Re (R(m))^{2}}\bigg(\frac{w(j+1,n)-w(j-1,n)}{2 \Delta x} \bigg)^{2} \nonumber \\ 
&+\frac{ \sigma_{hnf}}{\sigma_{f}} \frac{Ec M^{2}}{Re}(w(j,n))^{2} \bigg],  \label{c2-ftcst}
\end{align}
with initial and boundary conditions:
\begin{equation}
\label{c2-ftcsbc}
\begin{aligned}
&  w(j,1) = 0, \qquad \theta(j,1) = 0, \\
&   w(2,n) = w(1,n), \qquad  \theta(2,n) = \theta(1,n), \\
&w(J,n) = 0, \,\,\,\,\,\,\ \theta (J,n) = 1.
\end{aligned}
\end{equation}
The \textbf{stability of FTCS scheme} depends on the step size $\Delta x$ and $\Delta t$; for our simulations, we use 
$\Delta x=0.025$ and $\Delta t=0.0001$, which are suitable for satisfying the von Neumann stability condition.

\begin{figure}[h]
\begin{subfigure}{0.53\textwidth}
\includegraphics[width=1\linewidth, height=6cm]{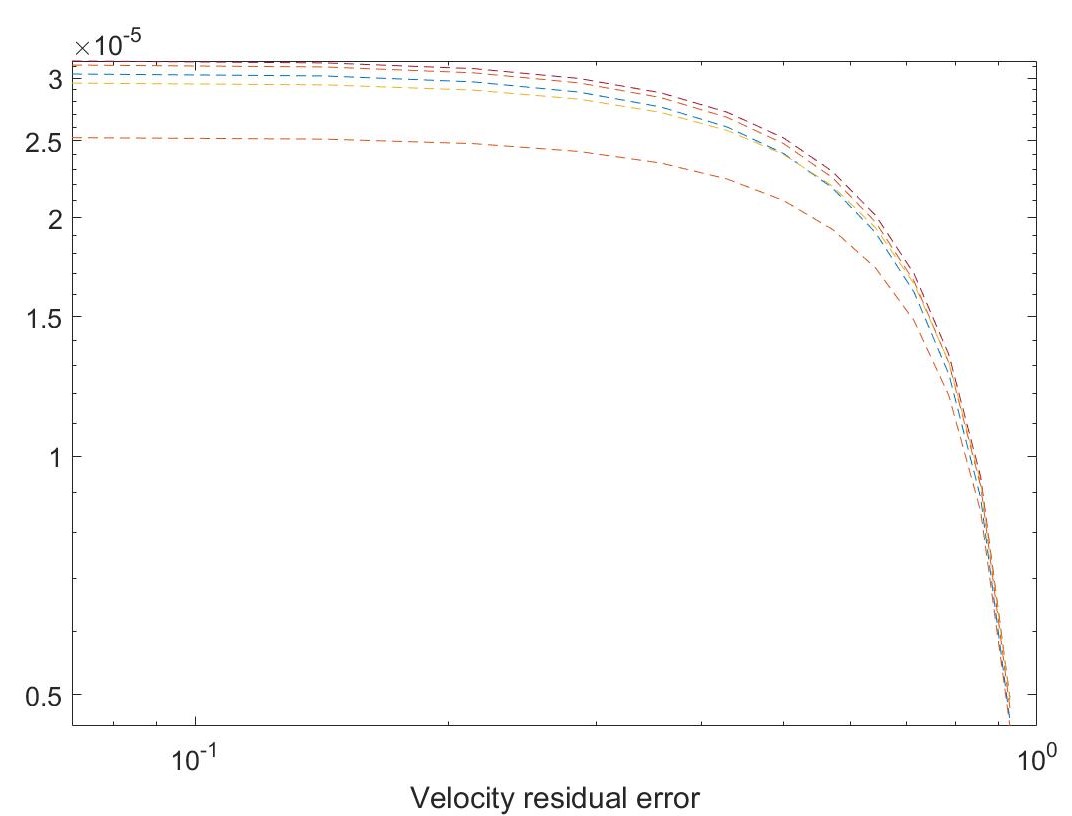} 
\caption{}
\end{subfigure} \hspace{-0.34em} \begin{subfigure}{0.53\textwidth}
\includegraphics[width=1\linewidth, height=6cm]{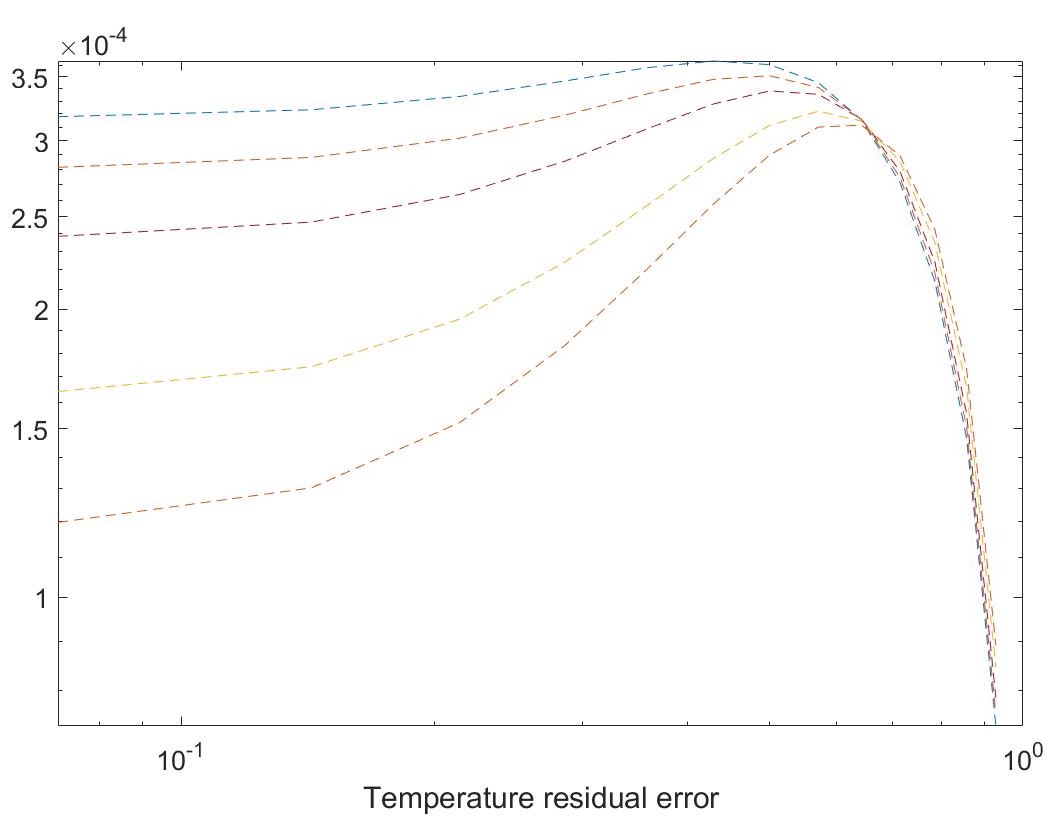}
\caption{}
\end{subfigure}
\caption{Convergence of momentum and energy profiles: Residual error analysis.}
\label{ResErrorUT}
\end{figure}

Figure~\ref{ResErrorUT} displays the residual error  for both velocity and temperature profiles maintaining $10^{-6}$ error across the entire domain. The initial residual errors are relatively high as significant corrections are made to the velocity and temperature field in the first few iterations. The residual error lines demonstrate a consistent and steady decay, indicating the stability of the numerical method. The values of emerging parameters occurring in the set of dimensionless PDEs are provided in Table \ref{valuesTable}.
\begin{table}[H]
\caption{Values of non-dimensional parameters in equations \eqref{c2-w1}-\eqref{c2-bc1}.} 
\label{valuesTable}
\begin{center}
\begin{tabular}{|c c|} 
\hline
Parameters & Fixed values \\[0.5ex]
Reynolds number $(Re)$ & 02.00 \\
Hartmann number $(M)$ & 00.50 \\ 
Casson fluid parameter $(\beta)$ & 00.50  \\ 
Eckert number $(Ec)$ & 04.00 \\
Volume fraction of Copper nanoparticles $(\phi_{1})$  & 00.01 \\
Volume fraction of Aluminium oxide nanoparticles $(\phi_{2})$  & 00.01 \\
Mean pressure gradient $(A_{0})$  & 01.41 \\
Prandtl number $(Pr)$  & 07.00 \\
Eccentricity $(e)$  & 00.50 \\
Angular frequency $(c_{1})$ & 01.00 \\
Hall current parameter $(m)$  & 01.50 \\
\hline
\end{tabular}
\end{center}
\end{table}
\section{Validation}
The validation of the finite difference technique is done by comparing the results with those obtained by other authors using the FTCS method. Tables \ref{table_3} and \ref{table_4} display axial velocity and temperature values of a hybrid nanofluid along the radial direction (denoted by $x$), juxtaposed with previously published findings from \cite{Tripathi, Algehyne,Basha}. The values from our study closely align with those in existing literature, indicating the reliability of our methodology. This comparison suggests that the behavior of the $Cu$-$Al_2O_3$-blood hybrid nanofluid axial velocity mirrors patterns observed in previous researches.
\begin{table}[H]
\caption{Comparison of axial velocity profiles for $Cu$-$Al_2O_3$-blood along the radius at $z = 0.71$ and $t = 1.15$: Present study versus  \cite{Tripathi} and \cite{Algehyne}.}
\label{table_3}
\begin{center}
\begin{tabular}{|c c c c|}
 \hline
$x$ & \cite{Tripathi} & \cite{Algehyne} & Present \\ [0.5ex] 
 \hline\hline
 0.0 & 0.5881 & 0.5890 & 0.5085 \\ 
 \hline
 0.1 & 0.5829 & 0.5828 & 0.5109 \\
 \hline
 0.2 & 0.5725 & 0.5727 & 0.5103 \\
 \hline
 0.3 & 0.5518 & 0.5539 & 0.4890 \\
 \hline
 0.4 & 0.5215 & 0.5259 & 0.4481 \\
 \hline
 0.5 & 0.4802 & 0.4865 & 0.4185 \\
 \hline
 0.6 & 0.4257 & 0.4334 & 0.3820 \\
 \hline
 0.7 & 0.3549 & 0.3608 & 0.2868 \\
 \hline
 0.8 & 0.2634 & 0.2689 & 0.2273 \\
 \hline
 0.9 & 0.1473 & 0.1481 & 0.0836 \\
 \hline
 1.0 & 0.0000 & 0.0000 & 0.0000 \\
 \hline
\end{tabular}
\end{center}
\end{table}
\begin{table}[H]
\caption{Comparison of temperature profiles for $Cu$-$Al_2O_3$-blood along the radius at $z = 0.71$ and $t = 1.15$: Present study versus  \cite{Basha}.}
\label{table_4}
\begin{center}
\begin{tabular}{|c c c|}
 \hline
$x$ & \cite{Basha} & Present \\ [0.5ex] 
 \hline\hline
 0.0 & 0.093989 & 0.0006628  \\ 
 \hline
 0.1 & 0.101855 & 0.000314  \\
 \hline
 0.2 & 0.126176 & 0.001481  \\
 \hline
 0.3 & 0.168902 & 0.003339  \\
 \hline
 0.4 & 0.232505 & 0.015540  \\
 \hline
 0.5 & 0.318937 & 0.032469  \\
 \hline
 0.6 & 0.428428 & 0.065463  \\
 \hline
 0.7 & 0.558456 & 0.224699  \\
 \hline
 0.8 & 0.703227 & 0.373057  \\
 \hline
 0.9 & 0.853959 & 0.792288  \\
 \hline
 1.0 & 1.000000 & 1.000000  \\
 \hline
\end{tabular}
\end{center}
\end{table}
\section{Discussion}
In this section, we discuss the influence of quantitative parameters emerging in the flow equations such as the volume fraction of $Cu$ and $Al_{2}O_{3}$ nanoparticles, Casson fluid parameter, stenosis height, Reynolds number, Eckert number and Hartmann number over momentum and temperature equation. For the computational part, $A_{0}=1.41$, $Re = 2.0$, $e = 0.5$, $c_{1}=1.0$, $\beta = 0.5$, $R=0.825$, $M=0.5$, $m=0.1$, $Pr=14$, $Ec=0.1$, $\phi_{1}=0.025$ and $\phi_{2}=0.025$ are taken into account. Quantitative assessment for axial velocity and temperature for the ordinary fluid, $Cu$-blood nanofluid, and $Cu$-$Al_2O_3$-blood hybrid nanofluid across the radial direction is stated in Table \ref{tcc}. The velocity decreases with increasing radial distance, approaching zero at the artery wall. This trend is consistent with the no-slip condition at the boundary, where the fluid has zero velocity relative to the wall. Meanwhile, the temperature increases with the radial distance from the center of the artery to the arterial wall. This increase indicates heat transfer from the arterial wall into the fluid, which is typically warmer at the wall due to the metabolic heat transfer from the surrounding tissues.
\begin{table}[H]
\centering
\caption{Table for radial distribution for the axial velocity and temperature of the ordinary fluid, $Cu$-blood, and $Cu$-$Al_2O_3$-blood along the radius.}
\begin{tabular}{|c|c|c|c|c|c|c|c|}
\hline
\multicolumn{1}{|c|}{$\frac{r}{R(z)}$} & \multicolumn{3}{c|}{Axial velocity} & \multicolumn{3}{c|}{Temperature} \\ \hline 
\multicolumn{1}{|c|}{} & Blood & $Cu$-blood & $Cu$-$Al_2O_3$-blood & Blood & $Cu$-blood & $Cu$-$Al_2O_3$-blood \\ \hline 
0.0 & 0.619304 & 0.515954 & 0.488037 & 0.000667 & 0.000530 & 0.000654 \\ \hline 
0.1 & 0.622360 & 0.518325 & 0.490295 & 0.000319 & 0.000258 & 0.000331 \\ \hline
0.2 & 0.610065 & 0.508765 & 0.481193 & 0.001572 & 0.001273 & 0.001535 \\ \hline
0.3 & 0.594437 & 0.496542 & 0.469561 & 0.003382 & 0.002890 & 0.003481 \\ \hline
0.4 & 0.542609 & 0.455489 & 0.430546 & 0.013923 & 0.013642 & 0.016396 \\ \hline
0.5 & 0.505500 & 0.425676 & 0.402252 & 0.028143 & 0.028920 & 0.034216 \\ \hline
0.6 & 0.460258 & 0.388930 & 0.367416 & 0.056496 & 0.059414 & 0.068576 \\ \hline
0.7 & 0.343539 & 0.292511 & 0.276153 & 0.203746 & 0.212628 & 0.230915 \\ \hline
0.8 & 0.271405 & 0.231995 & 0.218953 & 0.350292 & 0.359815 & 0.379618 \\ \hline
0.9 & 0.099306 & 0.085530 & 0.080677 & 0.785681 & 0.786146 & 0.794586 \\ \hline
1.0 & 0.000000 & 0.000000 & 0.000000 & 1.000000 & 1.000000 & 1.000000 \\ \hline
\end{tabular}
\label{tcc}
\end{table}
\begin{figure}[h]
\begin{subfigure}{0.53\textwidth}
\includegraphics[width=0.92\linewidth, height=6cm]{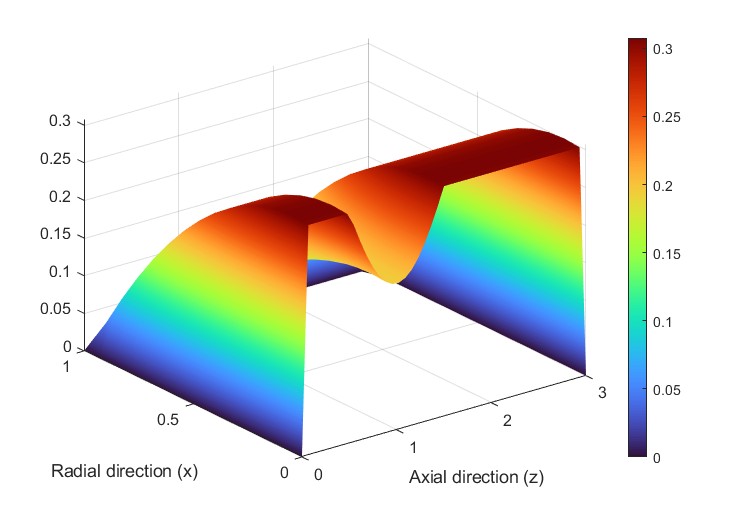} 
\caption{}
\label{surf_U}
\end{subfigure}\begin{subfigure}{0.53\textwidth}
\includegraphics[width=0.92\linewidth, height=6cm]{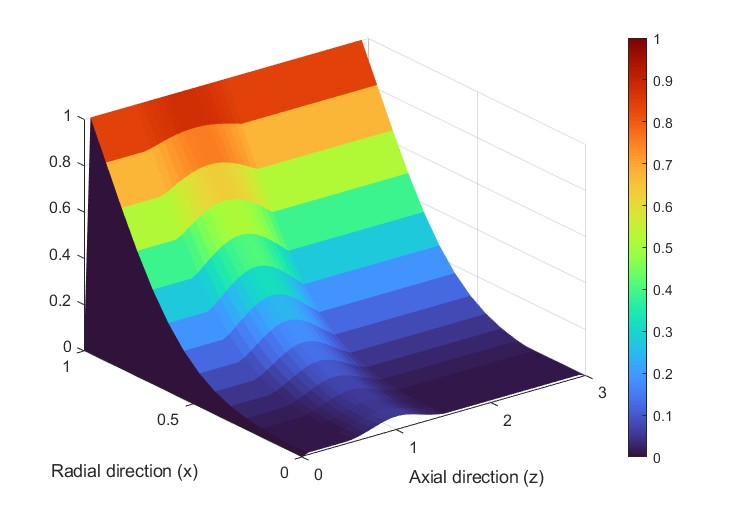}
\caption{}
\label{surf_T}
\end{subfigure}
\caption{Surface plot for the (a)velocity (b) and temperature distributions of hybrid nanofluid in a 36\% axisymmetric constricted stenotic artery.}
\label{surf_UT}
\end{figure}
Figure \ref{surf_UT} displays the 3D surface plot for axisymmetric blood velocity and temperature profiles in an elliptical stenotic arterial section with $x$ and $z$ as radial and axial directions, and the color range indicates the velocity and temperature spectrums. In Figure \ref{surf_U}, the stenosis is represented as a dip in the center of the plot, and the dark blue area of the artery axis indicates a zero velocity gradient. The yellow area at the artery walls represents the no-slip condition, where blood sticks due to viscosity. Stenosis causes acceleration and higher speeds at the edges of the constriction, followed by deceleration downstream. In Figure \ref{surf_T}, the blue region along the axis indicates a zero temperature gradient, implying constant temperature across the radial dimension due to symmetry. This means no net heat transfer across the flow perpendicular to it, consistent with axisymmetric flow conditions. Yellow regions at the artery periphery represent walls with $\theta(wall)=1$. The progression from blue at the center to yellow at the walls suggests a radial temperature gradient, with heat transferring from the wall toward the center of the artery.
\begin{figure}[H]
\centering
\begin{subfigure}{0.52 \textwidth}
\includegraphics[width=0.92\linewidth, height=6cm]{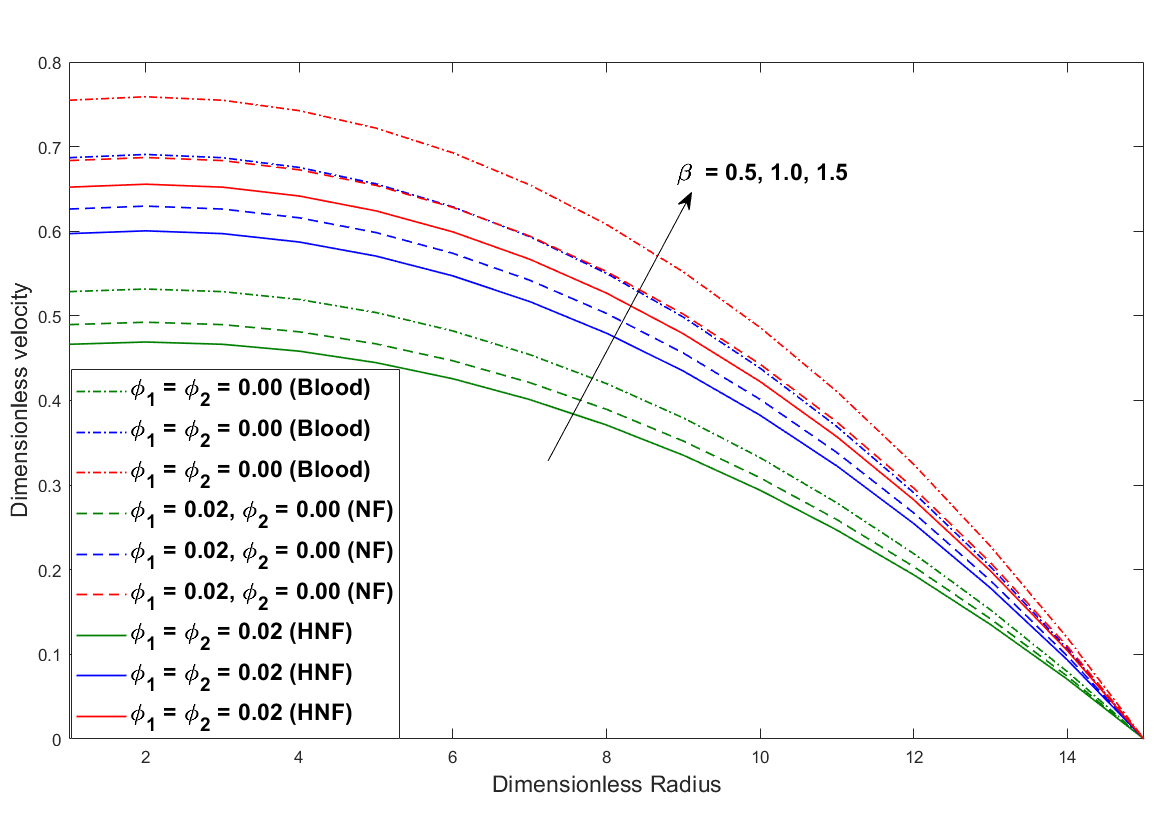} 
\caption{}
\label{casson_U}
\end{subfigure}\begin{subfigure}{0.52 \textwidth}
\includegraphics[width=0.92\linewidth, height=6cm]{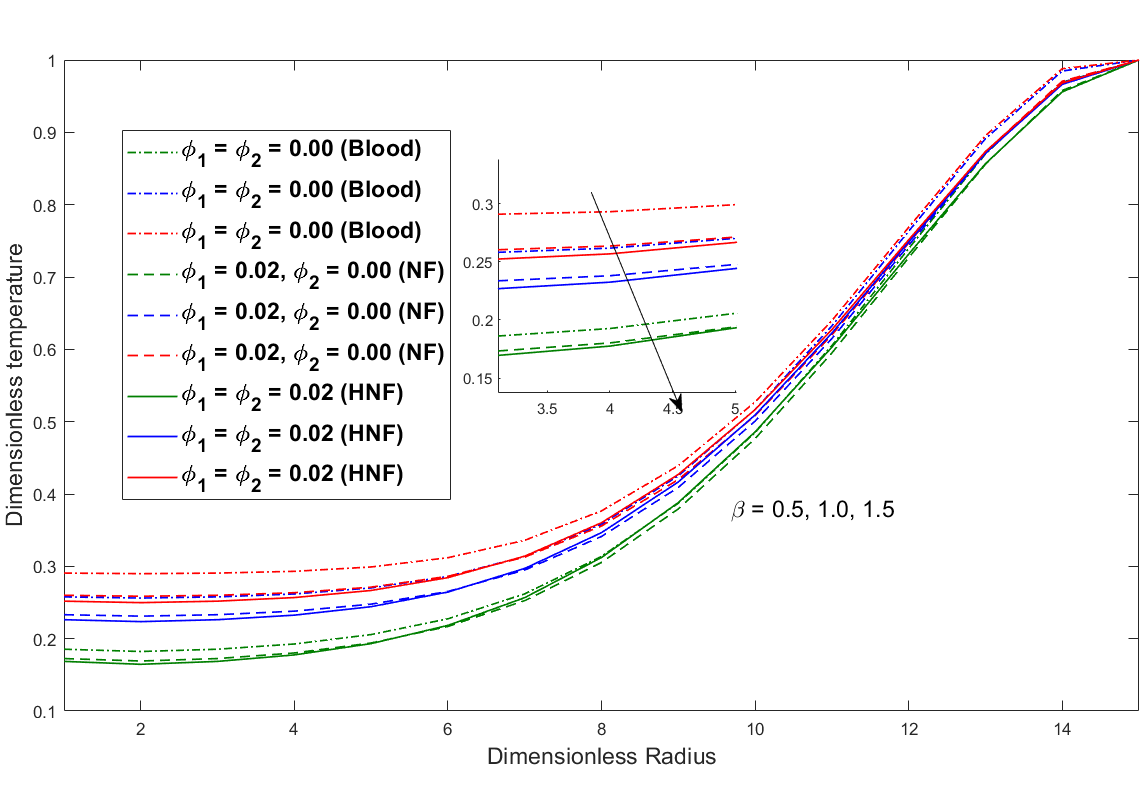}
\caption{}
\label{casson_T}
\end{subfigure}
\caption{Blood, $Cu$-blood and $Cu$-$Al_2O_3$-blood (a) velocity and (b) temperature profiles against variation in Casson Model Parameter $\beta$.}
\label{Casson_UT}
\end{figure}

A noticeable trend from Figure \ref{casson_U} is the escalation in velocity for blood without nanoparticles, ($\phi_1 = \phi_2 = 0.00$), the nanofluid ($\phi_1 = 0.02, \phi_2 = 0.00$) and the hybrid nanofluid ($\phi_1 = \phi_2 = 0.02$) with the rise in the Casson model parameter. The incremental increase in velocity with respect to $\beta$ across all mediums suggests a decrease in the effective viscosity of the blood, which allows for a more streamlined flow through the constricted region. Figure \ref{casson_T} shows that increasing the Casson model parameter $(\beta = 0.5, 1.0, 1.5)$ decreases the dimensionless temperature across all fluids. This indicates that when the blood exhibits more non-Newtonian properties and behaves like a Casson fluid, its capacity to retain heat increases. This is because reduced fluidity hinders convective heat transfer from the heated area.
\begin{figure}[H]
\centering
\begin{subfigure}{0.52 \textwidth}
\includegraphics[width=0.92\linewidth, height=6cm]{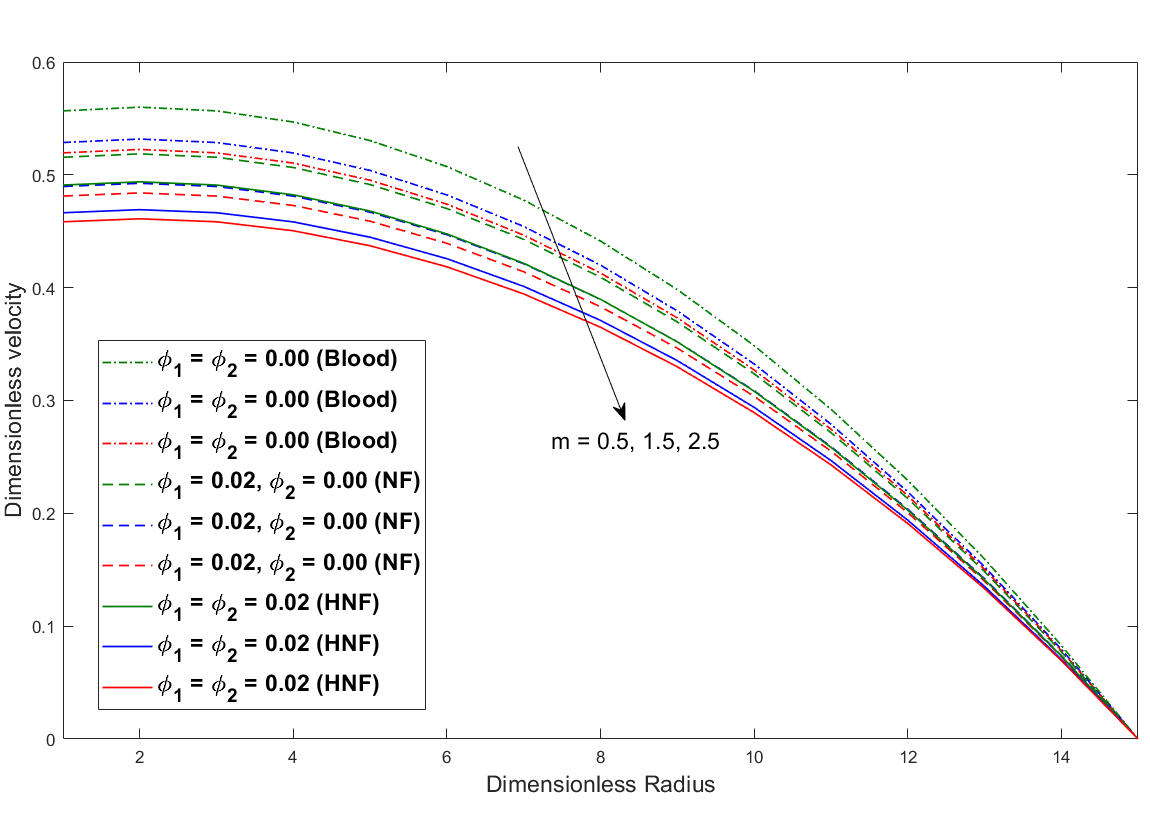} 
\caption{}
\label{Hall_U}
\end{subfigure}\begin{subfigure}{0.52 \textwidth}
\includegraphics[width=0.92\linewidth, height=6cm]{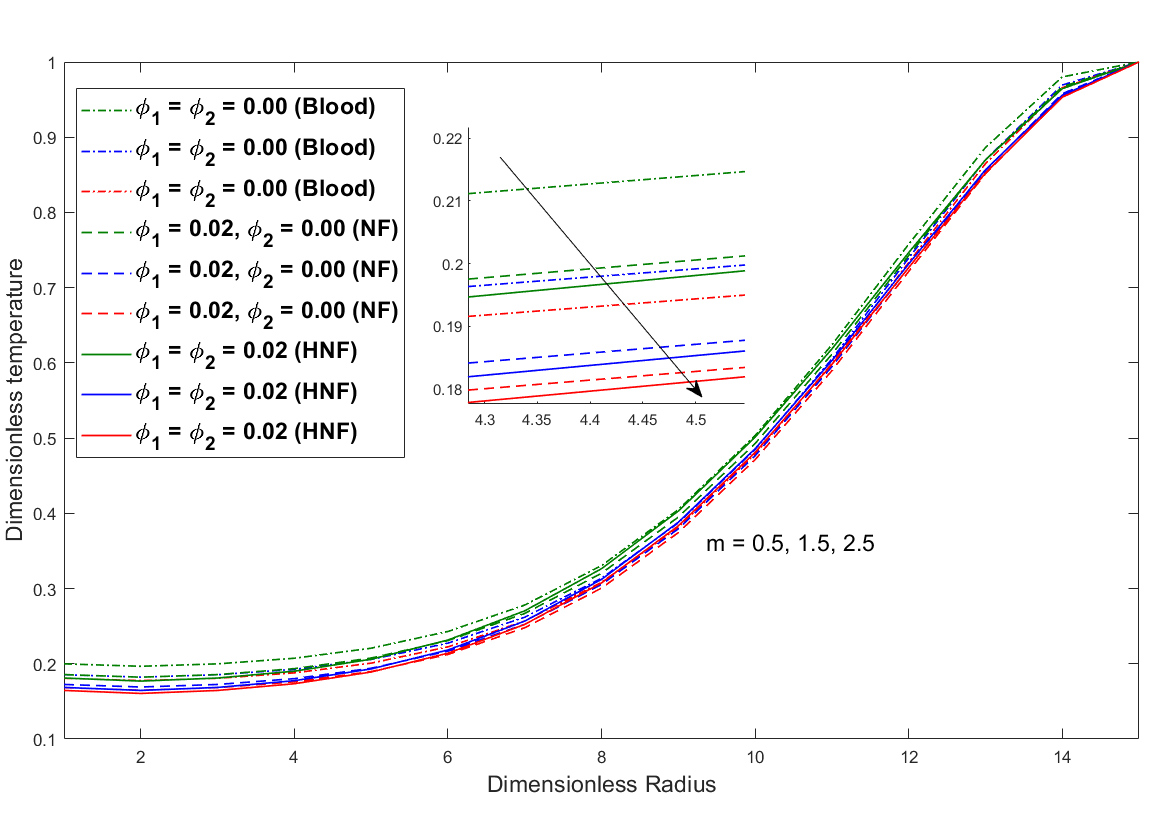}
\caption{}
\label{Hall_T}
\end{subfigure}
\caption{Blood, $Cu$-blood and $Cu$-$Al_2O_3$-blood (a) velocity and (b) temperature profiles against variation in Hall current parameter $m$.}
\label{Hall_UT}
\end{figure}
Non-Newtonian velocity and temperature profiles are illustrated in Figure \ref{Hall_UT} for ordinary fluid, nanofluid, and hybrid nanofluid against the variation in the strength of Hall current parameter $m$. With increasing values of $m$ in Fingure \ref{Hall_U}, a decrease in the dimensionless velocity is observed for the base fluid representing blood, $Cu$-blood, and $Cu$-$Al_2O_3$-Blood. This indicates that the Hall current acts to slow down the flow due to the interaction between the induced electric field and the charge carriers in the blood, which creates a force opposing the flow. From Figure \ref{Hall_T}, an increase in the Hall current parameter generated by the electromotive force $(EMF)$ transverse to current in conducting fluid corresponds to a decrease in the dimensionless temperature for the base fluid/blood, the nanofluid and the hybrid nanofluid. 
\begin{figure}[H]
\centering
\begin{subfigure}{0.52 \textwidth}
\includegraphics[width=0.92\linewidth, height=6cm]{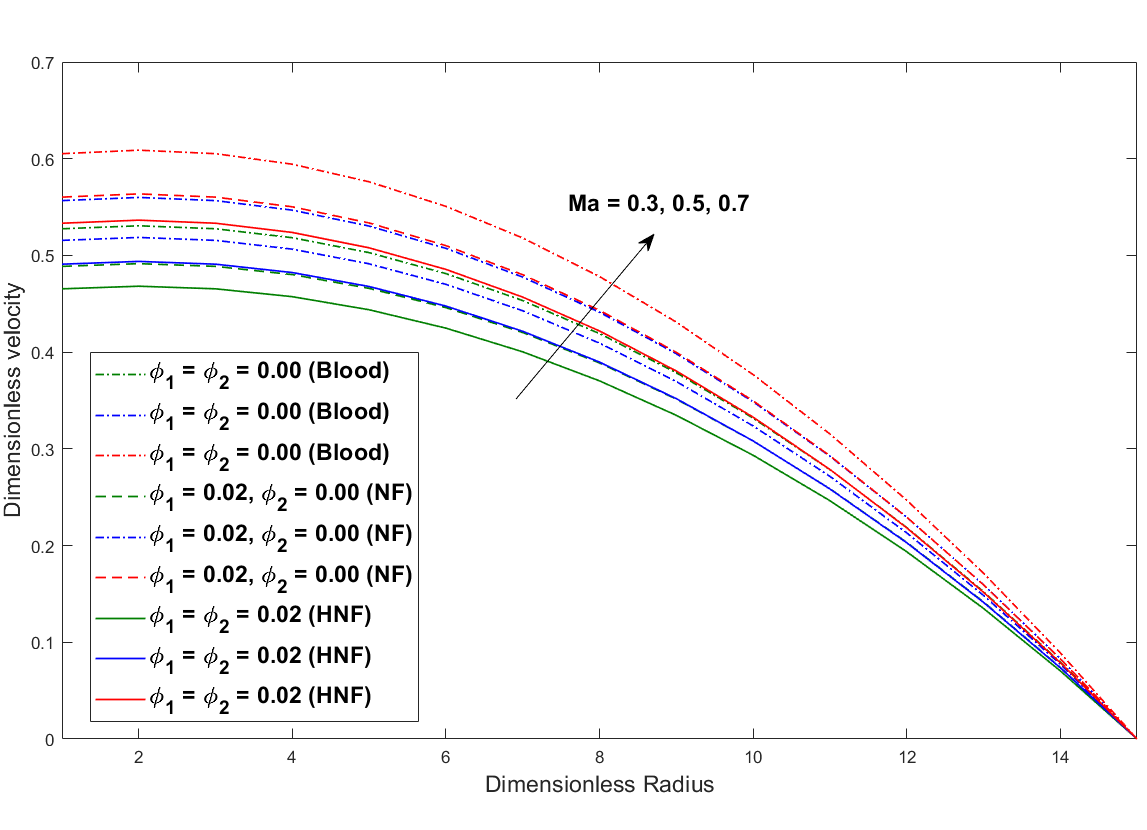} 
\caption{}
\label{Hartmann_U}
\end{subfigure}n\begin{subfigure}{0.52 \textwidth}
\includegraphics[width=0.92\linewidth, height=6cm]{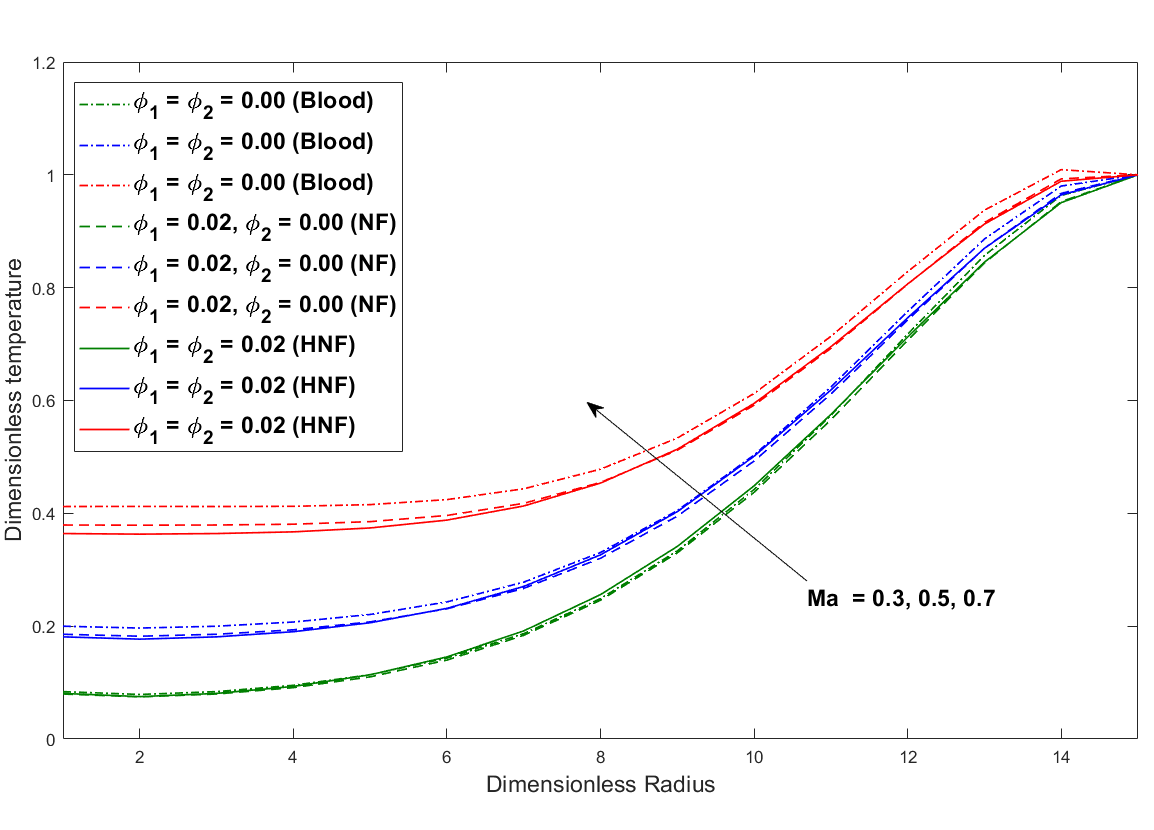}
\caption{}
\label{Hartmann_T}
\end{subfigure}
\caption{Blood, $Cu$-blood and $Cu$-$Al_2O_3$-blood (a) velocity and (b) temperature profiles against variation in dimensionless Hartmann number $Ma$.}
\label{Hartmann_UT}
\end{figure}

The magnetic parameter, or Hartmann number, is a dimensionless quantity in magnetohydrodynamics that characterizes the influence of a magnetic field on the flow of an electrically conducting fluid. As shown in Figure \ref{Hartmann_U}, there is an increased velocity with an increase in the Hartmann number $(Ma)$ across all fluids examined. The magnetic field imposes a force on the moving charged particles within the fluid, which aligns the flow and reduces the velocity gradients, leading to a more streamlined flow profile. The fluid temperature also increases in Figure \ref{Hartmann_T} when a magnetic field is applied to a conducting fluid, inducing a perpendicular Lorentz force that can decrease turbulence and lead to laminar flow. Laminar flow has less mixing and surface contact, reducing heat dissipation from the fluid and a higher temperature.
\begin{figure}[H]
\centering
\begin{subfigure}{0.52 \textwidth}
\includegraphics[width=0.92\linewidth, height=6cm]{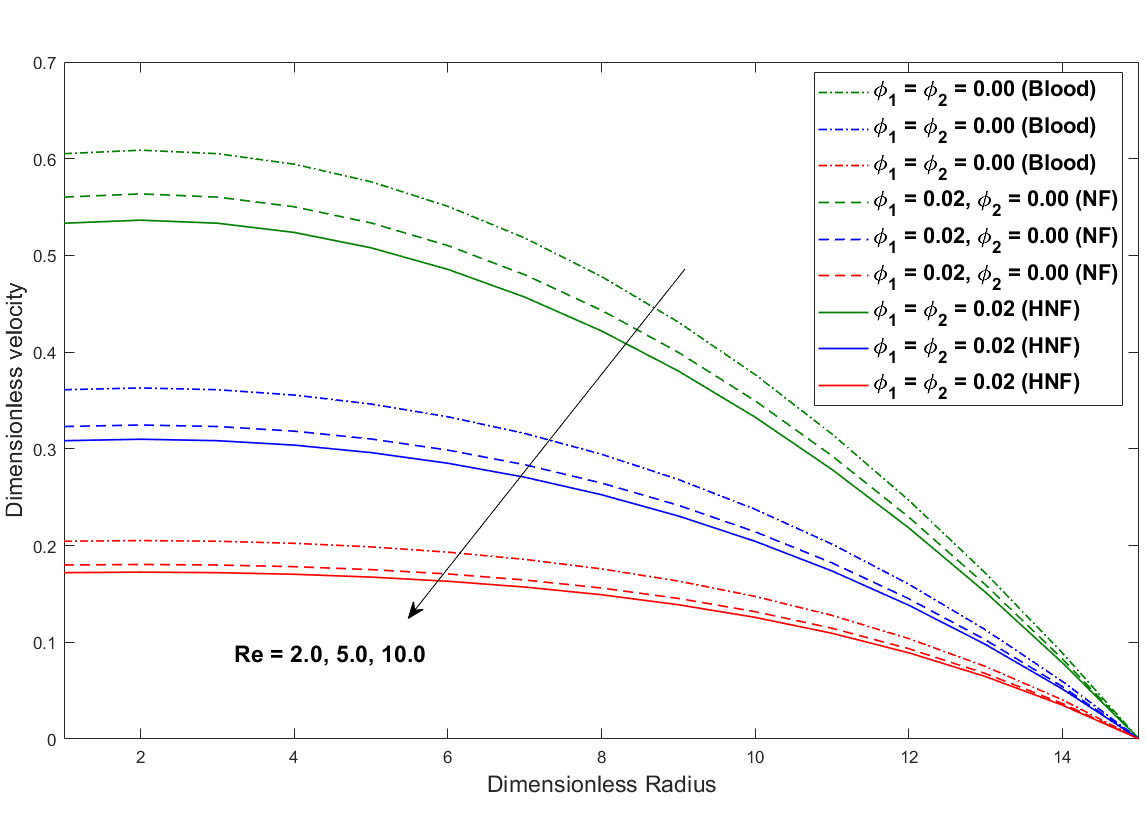} 
\caption{}
\label{Re_U}
\end{subfigure}\begin{subfigure}{0.52\textwidth}
\includegraphics[width=0.92\linewidth, height=6cm]{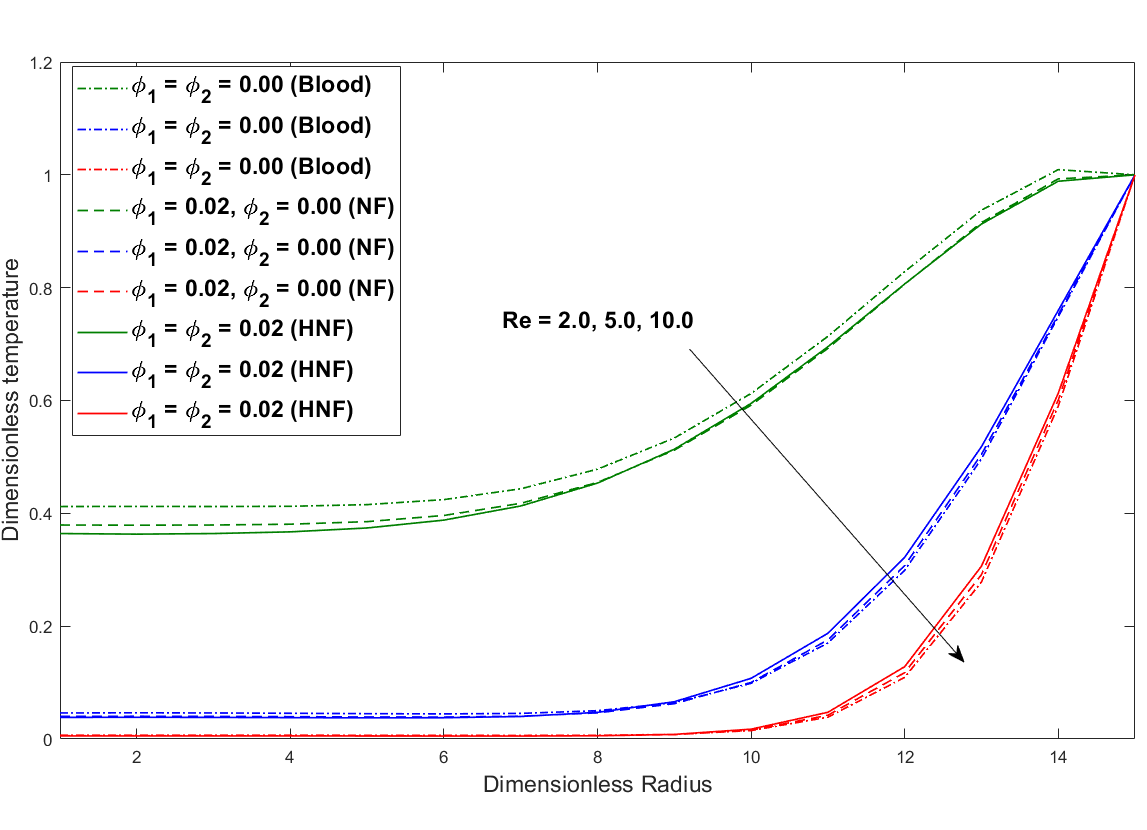}
\caption{}
\label{Re_T}
\end{subfigure}
\caption{Blood, $Cu$-blood and $Cu$-$Al_2O_3$-blood (a) velocity and (b) temperature profiles against variation in dimensionless Reynolds number $Re$.}
\label{Re_UT}
\end{figure}

From Figure \ref{Re_UT}, both momentum and temperature of fluid studied (Blood, $Cu$-blood and $Cu$-$Al_2O_3$-blood) tend to decline for rising Reynolds number. An increase in Reynolds number $(Re = 2.0, 5.0, 10.0)$ causes a reduction in viscous forces, which, in turn, decreases the velocity profiles in Figure \ref{Re_U}. The graph also illustrates that mass transfer is higher for all fluids. Similarly, the temperature profile in Figure \ref{Re_T} also decreases with an increase in the dimensionless Reynolds number for blood, $Cu$-blood and $Cu$-$Al_2O_3$-blood. This trend suggests that, for higher Reynolds number $(Re = 2.0, 5.0, 10.0)$, the convective heat transfer is enhanced, leading to more efficient heat dissipation from the fluid and, thus, a lower temperature. 
\begin{figure}[H]
\centering
\begin{subfigure}{0.52 \textwidth}
\includegraphics[width=0.92\linewidth, height=6cm]{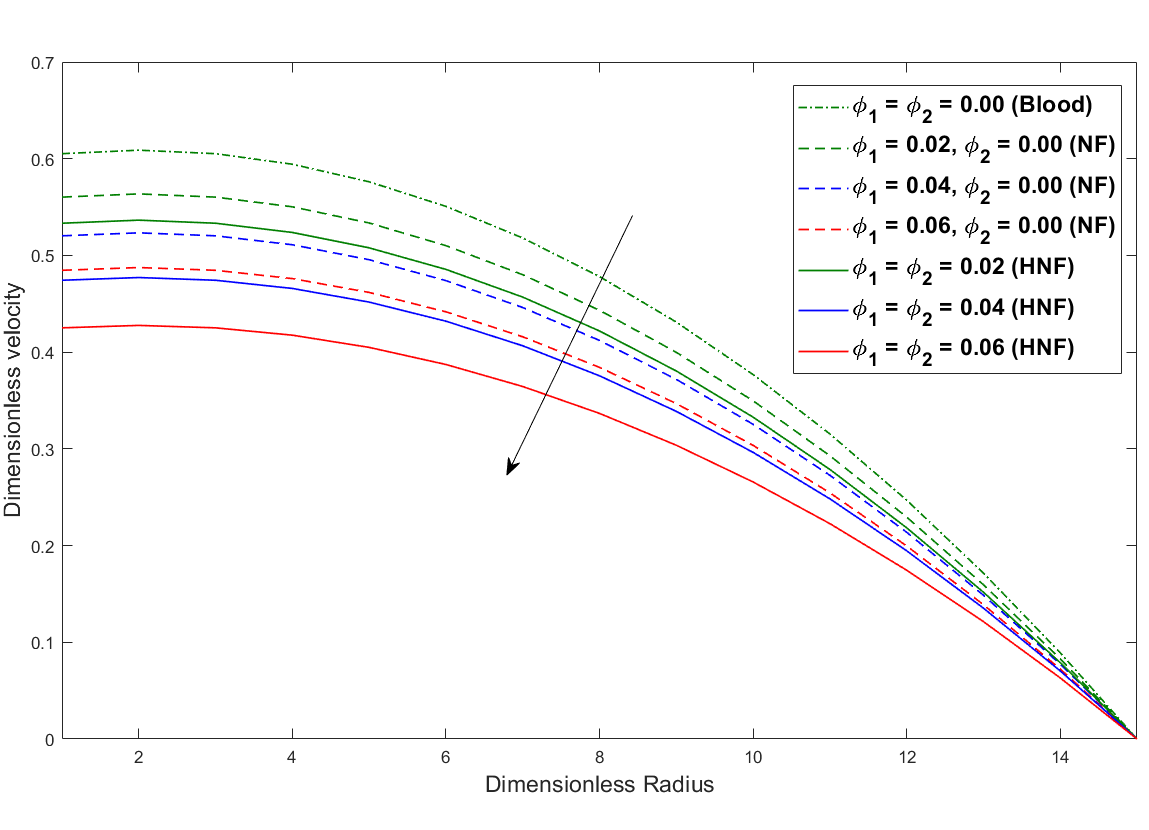} 
\caption{}
\label{NP_U}
\end{subfigure}\begin{subfigure}{0.52\textwidth}
\includegraphics[width=0.92\linewidth, height=6cm]{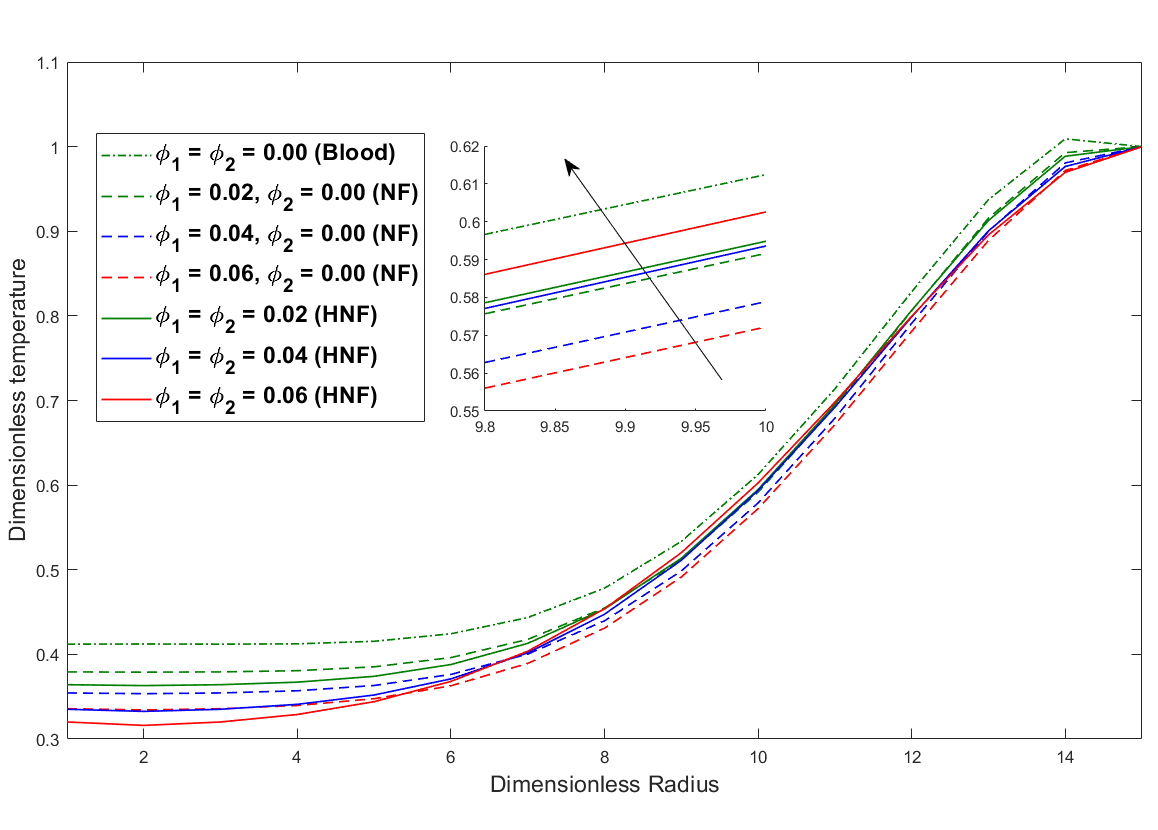}
\caption{}
\label{NP_T}
\end{subfigure}
\caption{Blood, $Cu$-blood and $Cu$-$Al_2O_3$-blood (a) velocity and (b) temperature profiles against variation in volume fraction $\phi_1, \phi_2$ for $Cu$ and $Al_2O_3$.}
\label{NP_UT}
\end{figure}
Opposite effects of an increase in volume fraction of copper and alumina nanoparticles on blood, nanofluid, and hybrid nanofluids velocity and temperature are observed in Figure \ref{NP_UT}. With increasing nanoparticle volume fractions $(\phi_1, \phi_2  = 0.02, 0.04, 0.06)$, a reduction in the velocity profile is observed in Figure \ref{NP_U}. Scientific literature has extensively studied the impact of introducing nanoparticles to blood, which increases its viscosity. Nanoparticles interact with blood plasma and cells, resulting in heightened resistance to flow, particularly at high concentrations. This effect leads to increased blood viscosity, reducing its mobility and thickness. The rise in temperature profiles in Figure \ref{NP_T} can be explained by the enhanced thermal conductivity that nanoparticles impart to the base fluid. It can be noticed that the hybrid nanofluid demonstrates more efficient heat transfer compared to the base fluid, likely due to the combined effect of $Cu$ and $Al_2O_3$ nanoparticle's thermal conductivities.\\
\begin{figure}[H]
\centering
\begin{subfigure}{0.52\textwidth}
\includegraphics[width=0.92\linewidth, height=6cm]{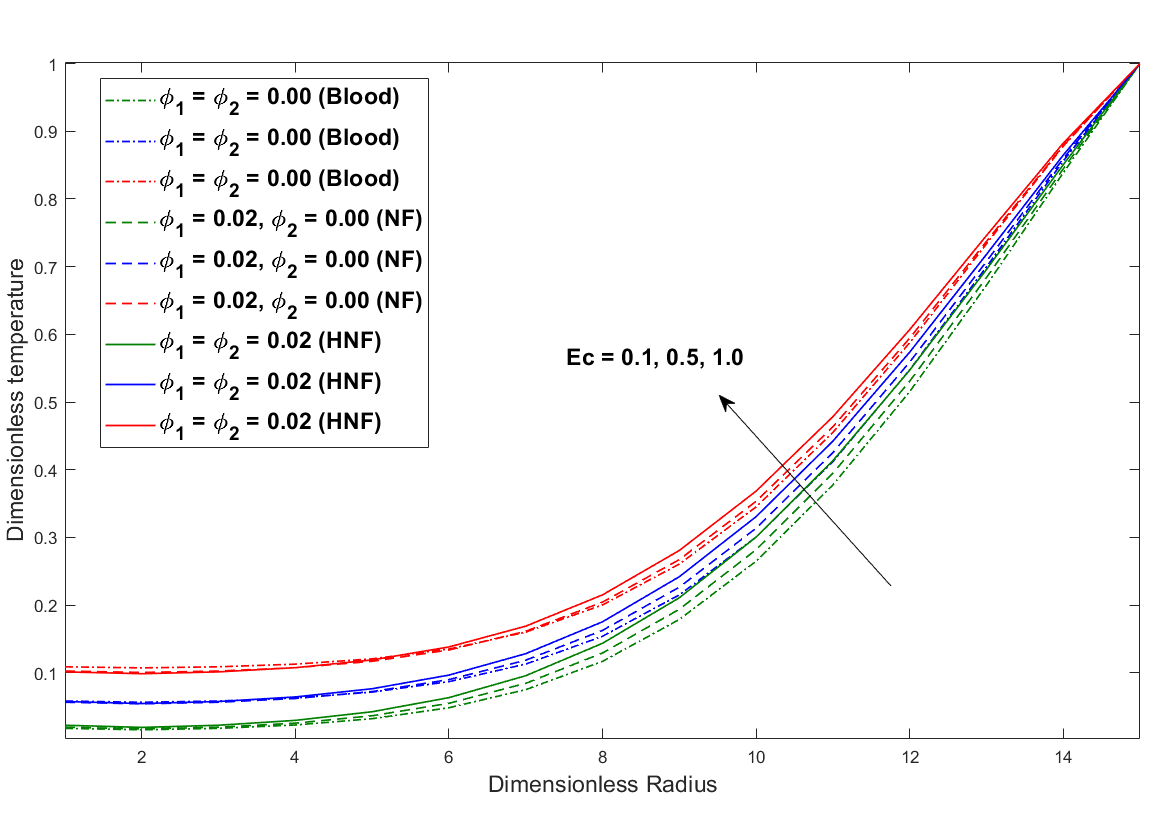} 
\caption{}
\label{Ec_T}
\end{subfigure}\begin{subfigure}{0.52\textwidth}
\includegraphics[width=0.92\linewidth, height=6cm]{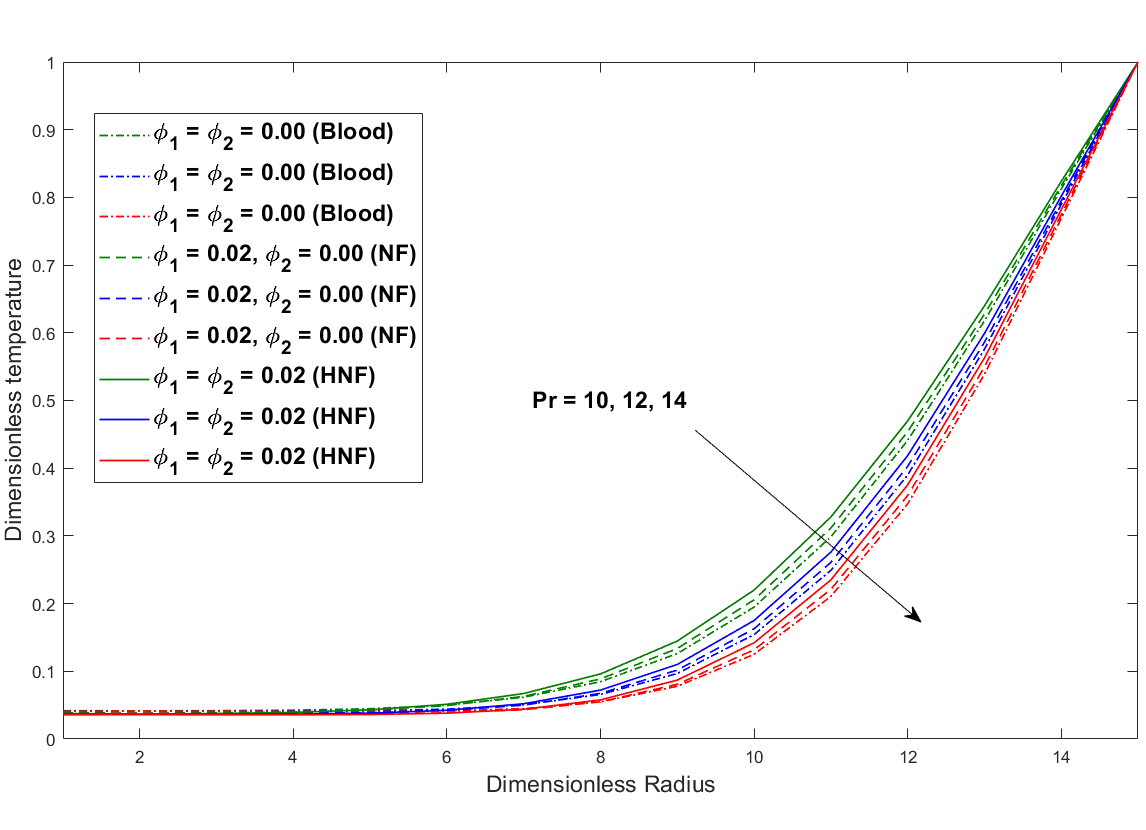}
\caption{}
\label{Pr_T}
\end{subfigure}
\caption{Blood, $Cu$-blood and $Cu$-$Al_2O_3$-blood temperature profiles against variation in dimensionless (a) Eckert number $Ec$ and (b) Prandtl number $Pr$.}
\label{EcPr_T}
\end{figure}
In Figure \ref{Ec_T}, there is a noticeable increase in temperature profile for the blood, nanofluid, and hybrid nanofluid for variation in dimensionless Eckert number $(Ec)$. This trend indicates viscous dissipation, where the fluid kinetic energy is converted into heat due to friction and viscous forces within the fluid. An increase in the Pr $(10.0, 12.0, 14.0)$ in Figure \ref{Pr_T} suggests that momentum diffusivity (viscosity) dominates over thermal diffusivity. As Pr increases, the thermal boundary layer becomes thicker relative to the velocity boundary layer, meaning heat is less efficiently diffused away from the heated surface, potentially leading to lower temperatures in the fluid bulk.\\
\begin{figure}[H]
\centering
\begin{subfigure}{0.52\textwidth}
\includegraphics[width=0.92\linewidth, height=6cm]{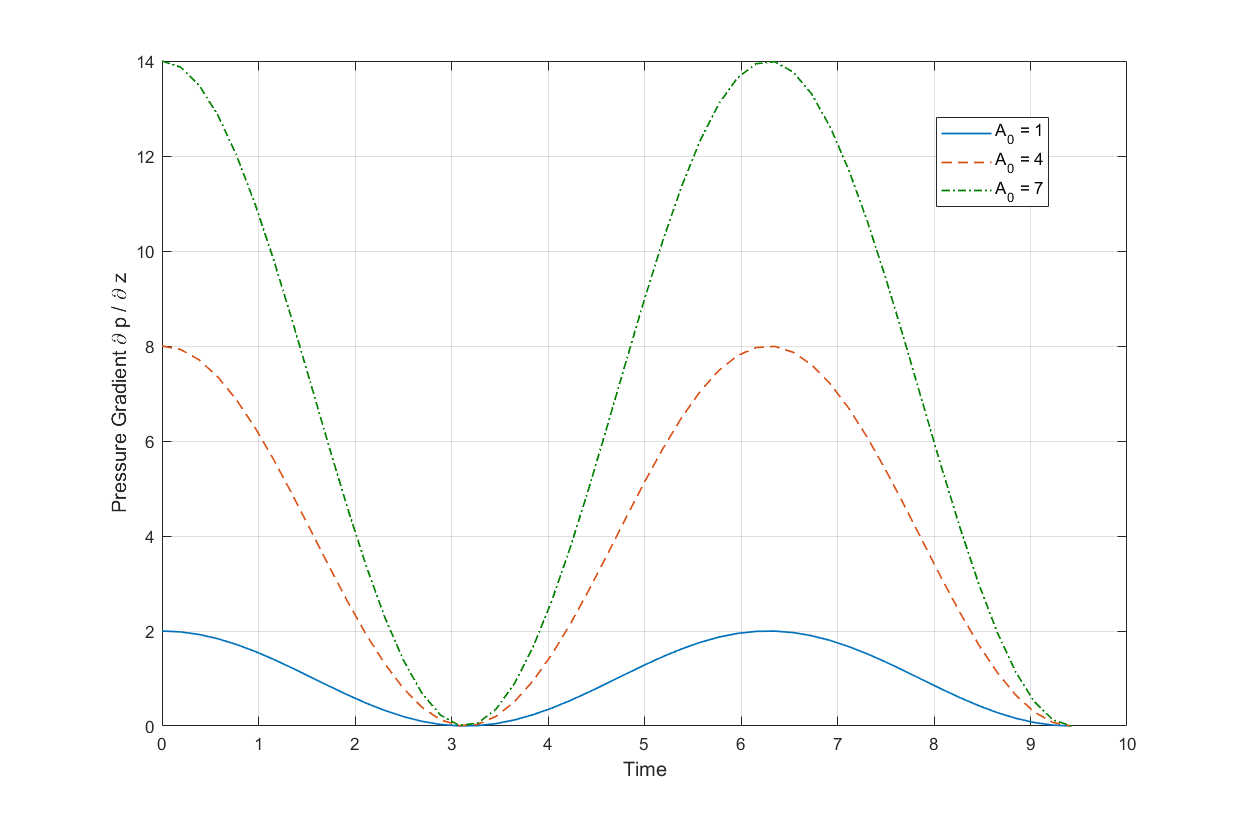} 
\caption{}
\label{p_grad}
\end{subfigure}\begin{subfigure}{0.52\textwidth}
\includegraphics[width=0.92\linewidth, height=6cm]{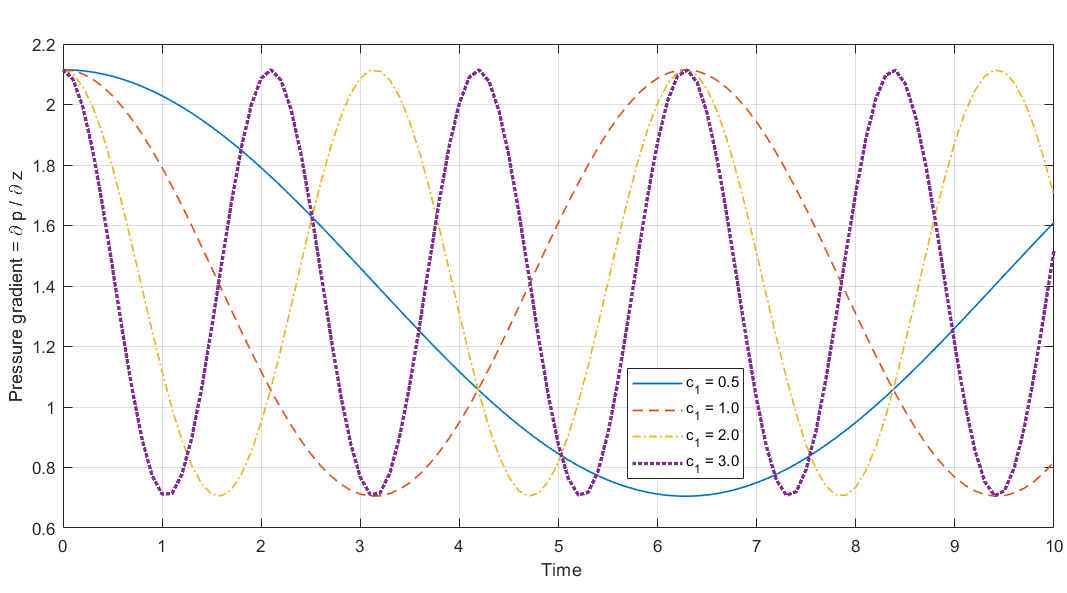}
\caption{}
\label{fig:c3-18}
\end{subfigure}
\caption{Pulsatile pressure gradient against variation in (a) average crossectional area $A_0$ and (b) frequency parameter $c_1$.}
\label{}
\end{figure}
Figure \ref{p_grad} illustrates the effect of varying the baseline amplitude $A_0$ on the pulsatile pressure gradient over time. $A_0=1$  represents a healthy individual with minimal arterial pressure oscillations indicating optimal cardiovascular function. The pressure gradient at $A_0=4$ shows more significant oscillations, typical of an average adult arterial pressure, where the baseline pressure is higher, resulting in more significant variations. $A_0=7$ indicates a condition like hypertension, where high pressures increase cardiovascular stress and risk. As shown in Figure \ref{fig:c3-18}, frequency parameter $c_1$ controls the speed at which the pressure gradient oscillates in response to heartbeats when the cardiovascular system is experiencing pulsatile blood flow. The pressure gradient oscillates at an increased rate as the $c_1$ parameter is increased, reflecting a more dynamic response to blood flow with pulsatile characteristics.\\
\begin{figure}[H]
\centering
\begin{subfigure}{0.54 \textwidth}
\includegraphics[width=0.95\linewidth, height=6.1cm]{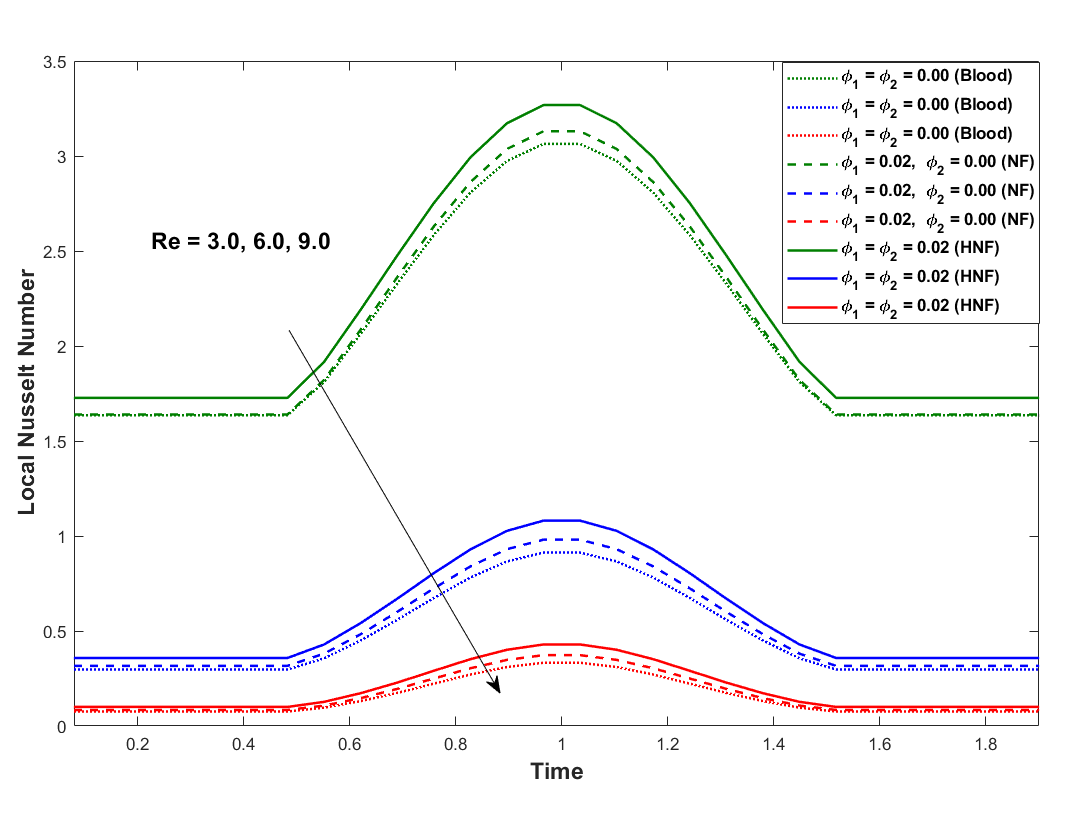} 
\caption{}
\label{Nu_Re}
\end{subfigure}\begin{subfigure}{0.54 \textwidth}
\includegraphics[width=0.95\linewidth, height=6.1cm]{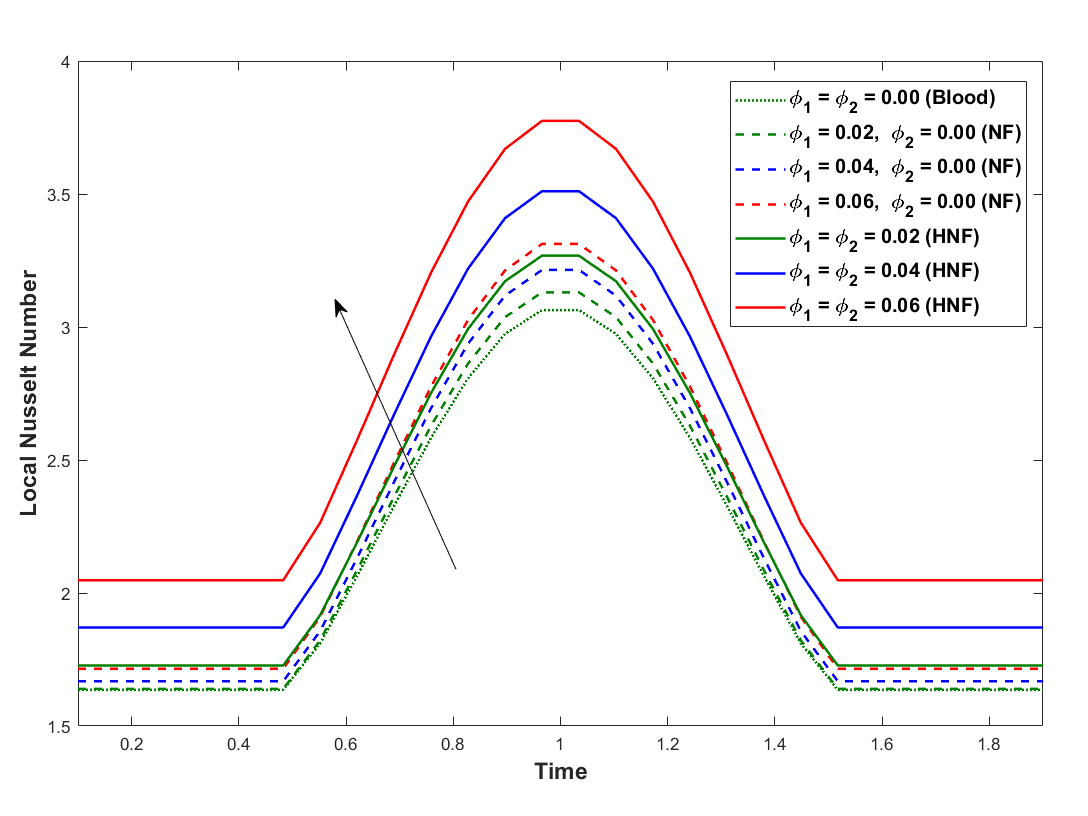}
\caption{}
\label{Nu_FNfHnf}
\end{subfigure}
\caption{Local Nusselt number response to varying (a) Reynolds number (b) concentrations of $Cu$ and $Cu$-$Al_2O_3$ nanoparticles for blood, nanofluid, and hybrid nanofluid.}
\label{Nu_Re_FNfHnf}
\end{figure}
Figure \ref{Nu_Re} illustrates how the heat transfer by fluid convection  improves as its Reynolds number increases due to increased mixing and thermal boundary layer disruption. The local Nusselt number, a dimensionless parameter, can measure this heat transfer efficiency.  
The increased volume fraction of added nanoparticles in Figure \ref{Nu_Re_FNfHnf} means more particles are available to facilitate heat transfer through mechanisms such as Brownian motion and thermophoresis, leading to an enhanced local Nusselt number. The graph indicates that the impact of hybrid nanoparticles on the local Nusselt number is more pronounced than that of single-component nanoparticles.\\ 
\begin{figure}[H]
\centering
\begin{subfigure}{0.54 \textwidth}
\includegraphics[width=0.95\linewidth, height=6.1cm]{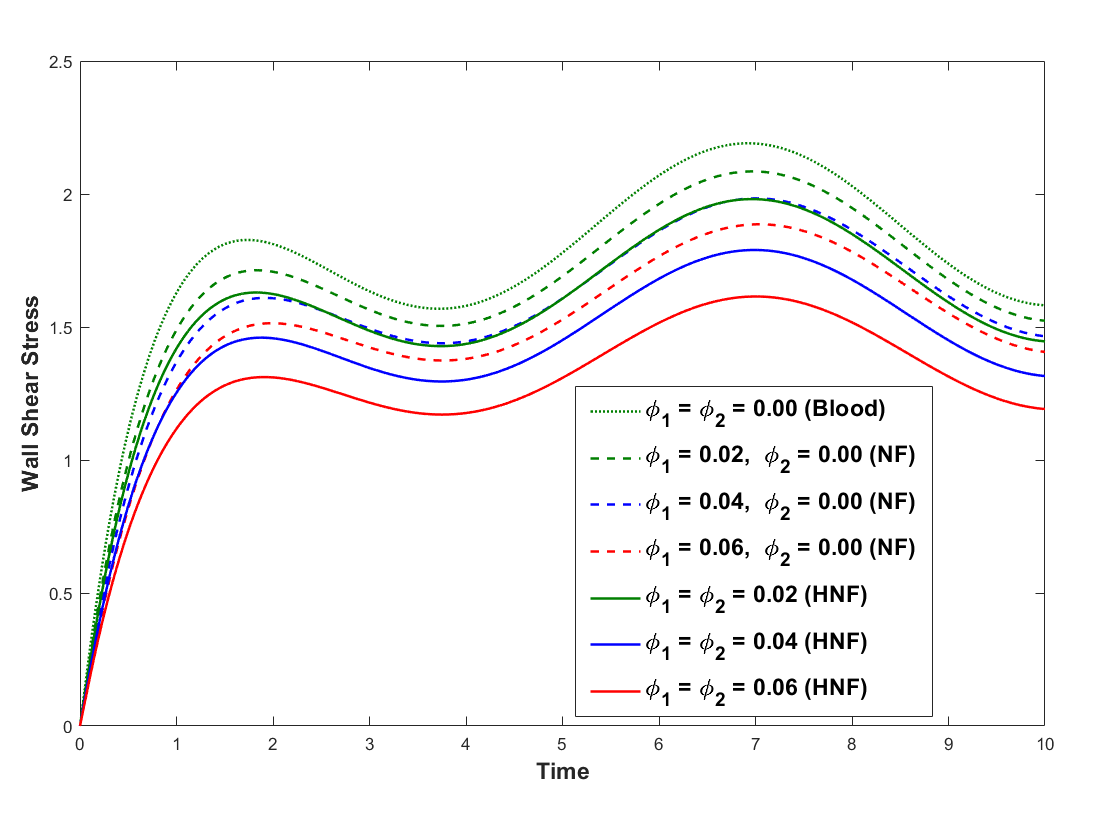} 
\caption{}
\label{wws_FNfHnf}
\end{subfigure}\begin{subfigure}{0.54 \textwidth}
\includegraphics[width=0.95\linewidth, height=6.1cm]{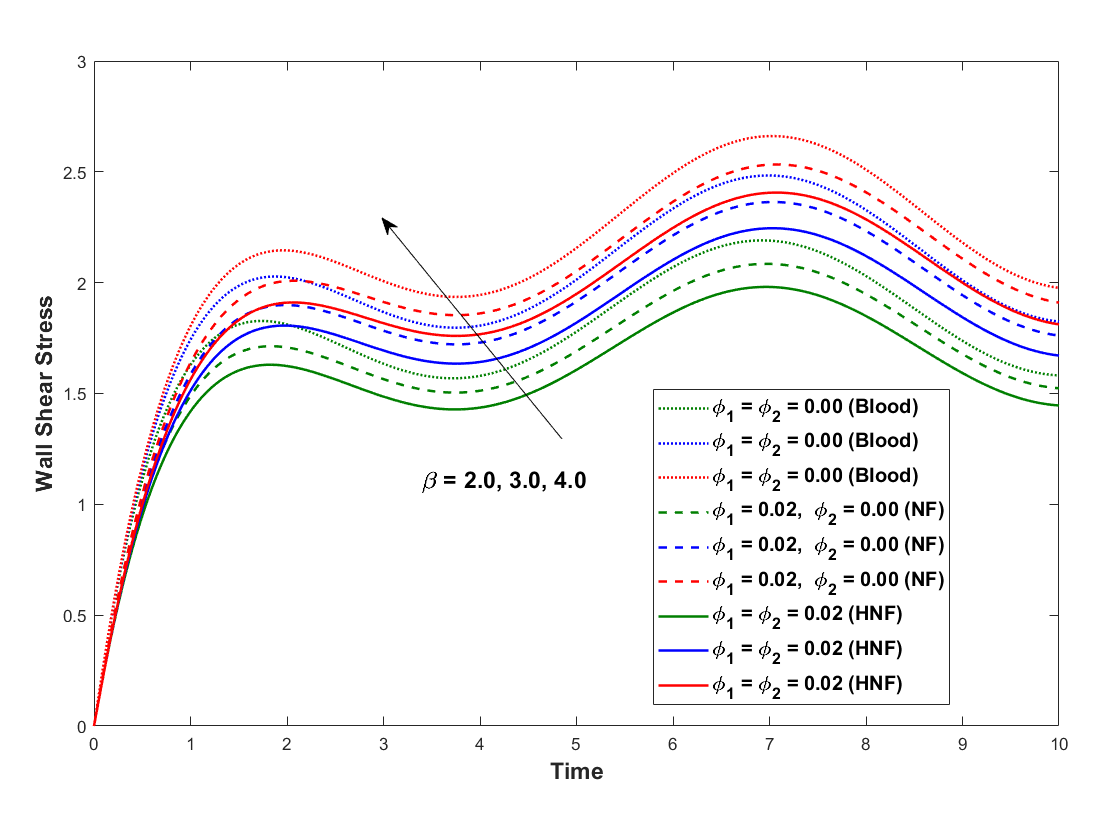}
\caption{}
\label{wws_casson}
\end{subfigure}
\caption{Wall shear stress response to varying (a) concentrations of $Cu$ and $Cu$-$Al_2O_3$ nanoparticles (b) Casson model parameter for blood, nanofluid, and hybrid nanofluid.}
\label{wws_B_FNfHnf}
\end{figure}
From Figure \ref{wws_FNfHnf}, reduction in wall shear stress is noticed for different values of $Cu$ and $Al_2O_3$ nanoparticles $(\phi_1, \phi_2 = 0.02, 0.04, 0.06)$. In combination with $Cu$ and $Al_2O_3$ nanoparticles, the effective viscosity and density of the fluid are further increased, which can significantly impact the velocity profile, especially in the boundary layer. The wall shear stress decreases as a result. 
For all types of fluids, including ordinary blood, nanofluids, and hybrid nanofluids, it is evident from Figure \ref{wws_casson} that the wall shear stress increases as the Casson model parameter is increased $(\beta = 2.0, 4.0, 6.0)$. In this case, $\beta$ indicates that the fluid's resistance to flow increases with $\beta$, requiring more stress to maintain the same flow rate.\\
\begin{figure}[H]
\centering
\begin{subfigure}{0.54\textwidth}
\includegraphics[width=0.95\linewidth, height=6cm]{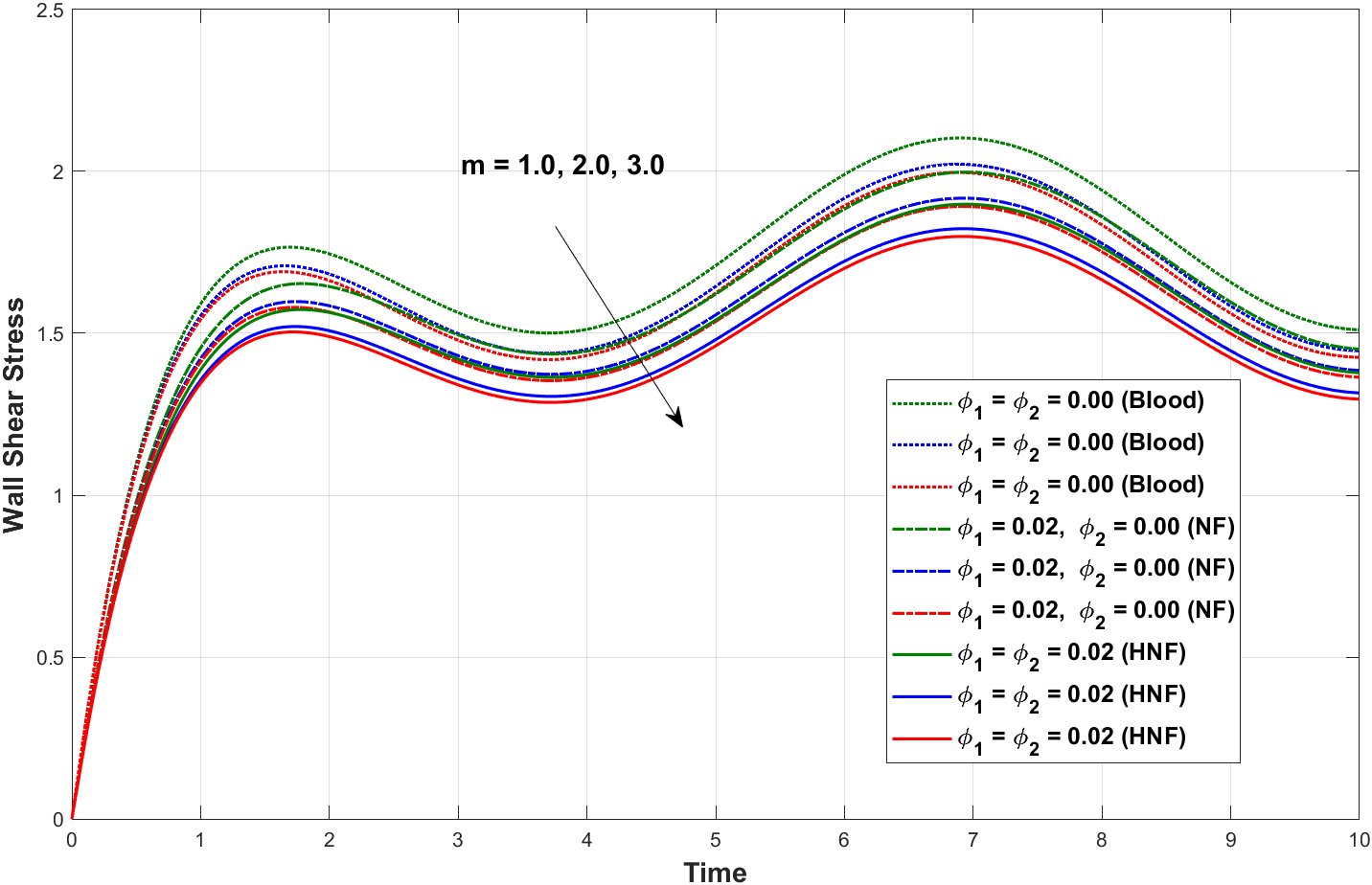} 
\caption{}
\label{wws_Hall}
\end{subfigure}\begin{subfigure}{0.59\textwidth}
\includegraphics[width=0.95\linewidth, height=6.2cm]{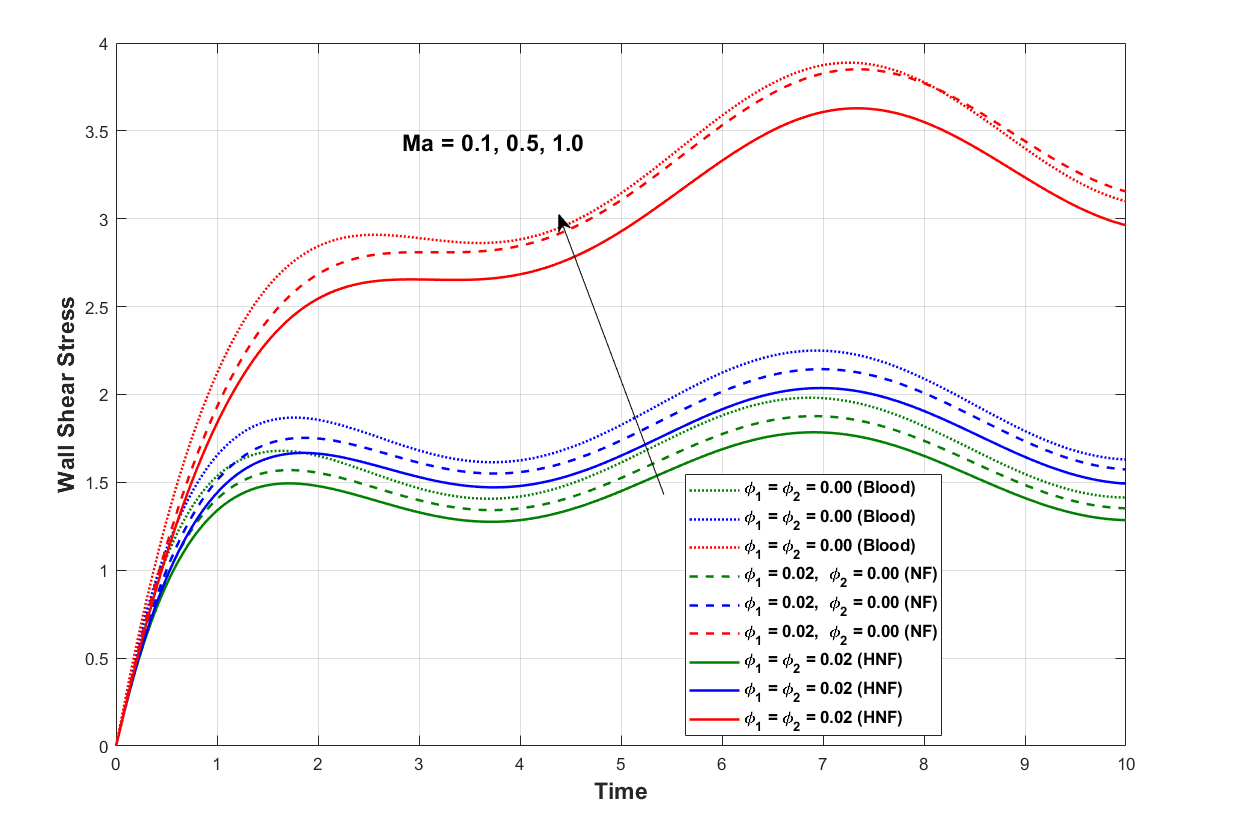}
\caption{}
\label{wws_Ma}
\end{subfigure}
\caption{Wall shear stress response to varying (a) Hall current parameter (b) Hartmann number for blood, nanofluid, and hybrid nanofluid.}
\label{wws_m_ma}
\end{figure}
Figure \ref{wws_Hall} shows that the wall shear stress for all fluid types decreases when the Hall current parameter $(m)$ increases from $1.0$ to $3.0$. It is evident from this decline that the magnetic field interacts with electrically charged particles in the flow, opposing its motion. Consequently, the wall shear stress could be reduced in the case of blood flow by reducing the velocity gradient at the vessel wall. The magnetic field interaction with charged particles in the fluid (Figure \ref{wws_Ma}) increases wall shear stress. The Lorentz force opposes the motion of a conducting fluid like blood when imposed with a magnetic field. Nanofluids have a higher electrical conductivity than conventional blood, resulting in a more significant impact by magnetic fields and, therefore, higher shear stresses. 
\section{Conclusions}
The investigation of unsteady axisymmetric hybrid nanofluid flow, comprising $Cu$-$Al_2O_3$-blood, through elliptical-shaped stenotic arteries has provided valuable insights into the complex dynamics of blood flow in the context of cardiovascular health. By utilizing the Casson fluid model to characterize the non-Newtonian behavior of blood, and considering factors such as pulsatile pressure gradients, magnetic fields, and nanoparticle additives, we have developed a comprehensive understanding of the multifaceted nature of flow phenomena within stenotic arteries. The set of partial differential equations defining the flow is simplified using the long-wavelength approximation. Numerical computations using MATLAB, stability analysys, and graphical studies of emerging parameters have enriched our understanding of the system behavior, revealing intricate relationships between various factors and flow characteristics. 

Key outcomes are:
\begin{enumerate}
\item The velocity distribution within the artery demonstrates a maximum along the axis, gradually diminishing towards the arterial walls. In contrast, the temperature profile exhibits the opposite behavior, with minimum values at the axis and increasing temperatures towards the walls.
\item An increase in the non-Newtonian Casson model parameter amplifies velocity while reducing the temperature profile. Antithetically, incorporating copper ($Cu$) and aluminum oxide ($Al_2O_3$) nanoparticles into the fluid results in a decrease in velocity and an increase in temperature.
\item Higher Hall current parameter and Reynolds number reduce velocity and temperature in hybrid nanofluid flow. Conversely, the magnetic field tends to enhance both velocity and temperature. 
\item Parameters such as the Eckert number contribute to improved thermal profiles, while the Prandtl number influences temperature reduction in hybrid nanofluid flow.
\item Local Nusselt number improves with increases in Reynolds number and volume fraction of nanoparticles. Increasing the volume fraction of nanoparticles, pulsatile frequency, and Hall current parameter decreases wall shear stress, while the Casson model and magnetic parameter increase wall shear rate.
\end{enumerate}

\bibliography{mybibfile}
\end{document}